%% file: 0-main.tex
\documentclass[preprint,11pt]{elsarticle}
\graphicspath{ {./figures/} }
\usepackage{hyperref}
\usepackage{float}
\usepackage{verbatim} 
\usepackage{apalike}

\usepackage{dsfont}
\usepackage{threeparttable}
\usepackage{indentfirst}
\usepackage{amsmath}
\usepackage{amsfonts}
\usepackage{color}
\usepackage{comment}
\usepackage{graphicx}
\usepackage{multirow}
\usepackage{algorithmic}
\usepackage{bm}
\usepackage{url}
\usepackage{booktabs}  
\usepackage{bbding}
\usepackage{enumitem}
\usepackage{float}
\usepackage{array}
\usepackage{caption}
\usepackage{float}
\usepackage{subfig}
\restylefloat{figure}
\restylefloat{table}

\journal{Expert Systems with Applications}

\biboptions{authoryear}

\begin{document}
\begin{frontmatter}

\title{Transition Information Enhanced Disentangled Graph Neural Networks for Session-based Recommendation}

\author[label1]{Ansong Li}
\ead{lasnling@gmail.com}

\cortext[cor1]{Corresponding author.}
\address[label1]{Department of Software Engineering, Xi'an Jiaotong University, 28 Xianning West Road, Xi'an, China}

\begin{abstract}
    Session-based recommendation (SBR) is a practical recommendation task that predicts the next item based on an anonymous behavior sequence, and its performance relies heavily on the transition information between adjacent items in the sequence. The current state-of-the-art methods in SBR employ graph neural network to model neighboring item transition information from global (i.e, other sessions) and local (i.e, current session) contexts. However, most existing methods treat neighbors from different sessions equally without considering that the neighbor items from different sessions may share similar features with the target item on different aspects and may have different contributions.  (e.g., \textit{color}, \textit{style}, \textit{size} of clothes). In other words, they have not explored finer-granularity transition information between items in the global context, leading to sub-optimal performance. In this paper, we fill this gap by proposing a novel method called Transition Information Enhanced Disentangled Graph Neural Network (TIE-DGNN) to capture finer-granular transition information between items and try to interpret the reason of the transition by modeling the various factors of the item. Specifically, we first propose a position-aware global graph at item-level, which utilizes the relative position information to distinguish the different types of neighbors, to model the neighboring item transition in the global context. Then, we slice item embeddings into blocks, each of which represents a factor, and use global-level disentangling layers equipped with position-aware embedding propagation to separately learn the factor embeddings over the global graph. Moreover, we employ distance correlation to encourage independence between each pair of factors. After obtaining the items of independent factor embeddings from the global context, we train local-level item embeddings by using attention mechanisms to capture transition information from the current session. Then, generating factor-aware inter-session embedding (global context) and intra-session embedding (local context) from two types of item embeddings, respectively. Finally, we use contrastive learning techniques to enhance the robustness of two types of session embeddings. To this end, our model considers two levels of transition information. Especially in global text, we not only consider finer-granularity transition information between items but also take user intents at factor-level into account to interpret the key reason for the transition. Extensive experiments on three real-world benchmark datasets demonstrate the superiority of our method over the SOTA methods.
\end{abstract}


\end{frontmatter}

\input{1-intro}

\input{2-rel-work}

\input{3-method}

\input{4-experiment}

\input{5-conclusion}

\bibliography{ref}

\end{document}

%% file: 1-intro.tex
\section{Introduction}
\label{introduction}
With the explosive growth of information on the Internet, recommendation systems are widely deployed on various platforms (e.g., web search, online shopping, etc.) to alleviate the data overload by recommending the desired content to users. Most existing typical recommended methods, such as collaborative filtering \citep{m-cf}\citep{NGCF}\citep{Light-GCN}, content-based methods \citep{CB-RS}\citep{F-CB} and trust-based methods \citep{tb}, utilize the user's identity information and long-term historical interactions to infer content that they are interested in. While these methods may fail in some real-world scenarios, such as unlogged-in users or those who have short-term interaction history. Thus, session-based recommendation, which aims to predict the next item based on an anonymous user's behavior sequence with chronological order, has attracted growing attention, and various methods have been proposed in this field. The earliest approaches to SBR (session-based recommendation) employ Markov chain \citep{FPMC} to predict the next user’s interest by modeling the sequential pattern in the session sequence. Due to its strong sequential assumption that the next item is solely based on the previous ones, it fails to capture long-term sequential dependence. 

To overcome the aforementioned problem,several methods apply deep learning techniques (i.e.,RNN-based or GNN-based) to model item-transition in the current session. Most RNN-based \citep{SR-RNN}\citep{P-RNN}\citep{I-RNN} methods treat session-based data as unidirectional sequences and model sequential patterns to capture item-transition information, which are then extended with attention network \citep{NARM} and memory network \citep{STAMP}. GNN-based methods convert session sequence into the graph and learn the transition relationship between distant items via item embedding propagating and updating over the graph. Wu et al.\citep{SR-GNN} is the first work to employ gated graph neural network to learn the item embedding in the session graph and achieves great success. Motivated by its success, several variants have been proposed \citep{TAGNN}\citep{GC-SAN}. All the above methods focus on the current session when modeling transition information, during which the performance is vulnerable to user behavioral sparsity and noisy data. To alleviate the problem, some methods \citep{CSRM}\citep{CoSAN} try to utilize collaborative information from other sessions for enhancing the performance of recommendation task. GCE-GNN \citep{GCE-GNN} traverses all sessions to find the most relevant neighbors and takes them as the unified neighbor set of each item as global information. Then it is combined with the local context (current session) to model the transition information. Compared with the previous methods, it achieves the best result.

Despite the progress achieved, we argue that the method of constructing global context information by unifying all filtered neighbors (from all sessions) into a set without distinguishing the different types of neighbors is unconsidered. The task of session-based recommendation is to predict the next item most likely to interact with the current session. Thus, effectively modeling the transition between neighboring items is crucial to enhance the accuracy of recommendations. While a uniform neighbor set derived from the global context can only enrich the current item representation via aggregation operations and cannot make the model obtain the transition information between them easily. This is because a unified neighbor set has no latent information to make the model obtain the common transition relationships between neighbors, such as whether the transition is from $A$ to $B$ or $B$ to $A$ or both, in most cases. Although these methods \citep{GCE-GNN}\citep{DGTN} estimate the transition information between items via focusing on the structure of the current session, the length of most sessions is very short, making it difficult to distill valuable information. Therefore, modeling the finer-grained transition relationship between items in the global context is crucial to improve the performance of the recommendation task.

In addition to building the transition relationship between neighboring items in the global context, more importantly, we need to further investigate the main reason for transition. The prediction in SBR made by matching the user's main intent of the session (which is represented by the session embedding) with candidate items. User's preference on items driven by various factors that characterize the item (e.g., \textit{phone} includes \textit{color}, \textit{resolution}, \textit{memory}), and more importantly, the preference could be dynamically changed along with the clicked items in a sequence. Thus, inferring the key factor that the user most cares about is crucial to interpret the transition relationship between items and enhance the performance of the recommendation. This is not trivial due to the fact that most existing embedding functions for SBR represent the item as a holistic representation (embedding), which does not distinguish the features of different features for an item. Modeling various factors on an item and accurately capturing the ones that users pay attention to remains a challenge.

In this paper, we propose a novel Transition Information Enhanced Disentangled Graph Neural Network (TIE-DGNN) to tackle the above problems. Specifically, we take global and local approaches to learn item embedding, respectively. For the global context (i.e., all the sessions), we first construct the position-aware global graph at item-level, which utilizes the relative position information to classify neighbors into different types so as to model the neighboring transition information. To characterize the factors of items, we slice item embeddings into multiple chunks with the assumption that each chunk represents a latent factor. Then, we employ global-level disentangling layers equipped with position-aware embedding propagation to separately learn the factor embeddings over the global graph and use distance correlation to encourage the independence factor-by-factor. By doing this, we can obtain independent factor embeddings that include transition information from global context to represent each item in our model. For local context (i.e., current session), we learn local-level (session-level) item embeddings by employing the attention mechanism to model neighboring transitions within the current session. After obtaining two types of embeddings for each item, we generate factor-aware inter-session embedding (global-level) and intra-session embedding (local-level) respectively by aggregating each item in the current session with attention weight, and then use contrastive learning techniques to enhance their robustness. To this end, we obtain two types of session embeddings by considering transition information from global and local contexts, respectively. Finally, we linearly combine these two session embeddings and make a prediction by calculating the similarity with the target item. 

To summarize, the main contributions of our work are threefold:
\begin{itemize}[leftmargin=1.5em]
   \item  We emphasize the importance of modeling transition information between neighboring items from global and local contexts. To the best of our knowledge, we are the first to consider fine-grained transition information modeling in the global context.
   \item We propose a novel transition information enhanced disentangled graph neural network for SBR. It can leverage the relative position information in the global context to capture the transition information between neighboring items, as well as represent the item with disentangled representations of factors to infer the main reason for the transition.
   \item We have conducted extensive experiments on three real-world benchmark datasets to demonstrate the effectiveness of our TIE-DGNN. Experimental results show that our proposed model outperforms the state-of-the-art methods. The ablation study further demonstrates the validity of different components in our model\footnote{The implementation of our model is available via https://github.com/AnsongLi/TIE-DGNN.}.

\end{itemize}

%% file: 2-rel-work.tex
\section{Related Work}
\subsection{Session-based Recommendation}
\textbf{Markov Chain for SBR.} The early methods for studying session-based recommendation are mainly based on Markov chain. Shani et al. \citep{MDP} employ markov decision processes (MDP) with appropriate initialization to capture the transition relationship between items. Rendle et al. \citep{FPMC} combine first-order markov chain and matrix factorization to capture sequential behavior between adjacent items in the session sequence and the general taste of a user, respectively.

\textbf{Deep Learning Methods for SBR.} With the boom of deep learning, many methods based on it apply to session-based recommendation. Hidasi et al. \citep{SR-RNN} employ the recurrent neural network called GRU4REC to model the sequential transition relationship between items by adopting a multi-layer Gated Recurrent Unit (GRU). Then they extend the model \citep{SR-RNN} with the parallel architecture \citep{P-RNN}. While Tan et al. \citep{I-RNN} enhance the model \citep{SR-RNN} by using data augmentation. In addition, Li et al. \citep{NARM} propose a hybrid encoder with an attention mechanism to extract the main purpose from the current session sequence. Liu et al. \citep{STAMP} emphasize the long-term and short-term interests of the current session, which are obtained by multi-layer perceptrons and attention mechanisms. MCPRN \citep{MCPRN} employs PSRU (GRU variant) in each mixture-channel to capture the multi-purpose of the current session. However, both RNN-based and Markov chain-based approaches have the strong sequential hypotheses, which means that they cannot capture transition relationships between distant items.

Recently, graph neural network has achieved great success in various fields. Thus, many methods based on it are proposed for session-based recommendation. Wu et al. \citep{SR-GNN} is the first work to convert session sequence into a graph to model high-order transition relationships between items and apply the gated graph neural network to learn item embeddings. Compared with RNN-based and Markov chain-based methods, the performance of SR-GNN improves by a large margin. Following the success of SR-GNN \citep{SR-GNN}, many variants have been proposed, such as GC-SAN \citep{GC-SAN} combines self-attention mechanism with graph neural network and TAGNN \citep{TAGNN} considers candidate item factors into recommendation task. Qiu et al.\citep{FGNN} propose the WGAT layer to serve as the item feature encoder that learns representation to assign different weights to different neighbors.

\textbf{Global Methods for SBR.} All the above methods only focus on the transition relationship between items in the current session. There are also works considering to leveraging the global information. Some collaborative filtering (CF) based methods \citep{CSRM}\citep{CoSAN} explore the the latest $n$ neighborhood sessions of the current session to model the global information at the session-level. But these methods may suffer from the problem of noise when integrating other sessions’ embeddings into the current one. Thus, some methods propose constructing a global graph to capture global information at the item-level. Qiu et al.\citep{FGNN-BCS} propose a broadly connected session (BCS) graph to link different sessions and a novel mask-readout function to improve session embedding. Zheng et al. \citep{DGTN} construct two channels to generate inter-session and intra-session embeddings from local and global contexts, respectively. Xia et al.\citep{DHCN} combine hypergraph model with self-supervised task to capture transition relationship between items from global context. Wang et al.\citep{GCE-GNN} traverse all sessions to find the most relevant neighbors and takes them as a unified neighbor set for each item as global information. Then, they combine it with the local context (current sesssion) to model the transition information. All these methods for modeling transition information from the global context are coarse-grained, which makes the models less explanatory. Thus, we propose a new method of constructing global graph by considering relative position information between items from all sessions, which captures the fine-grained transition relationship from the global context.      
\subsection{Disentangled Representation Learning}
Disentangled representation learning, which aims to learn independent factors behind the data, is quickly applied to various fields, such as computer vision\citep{Infogan}, text\citep{Text} and topic modeling \citep{topic}. In the field of recommendation, Ma et al.\citep{Macrid-VAE} is the first work to learn disentangled representations based on user behaviors. With the booming of graph neural network, Ma et al.\citep{D-GCN} applyed disentangled representation learning in graph convolutional network to achieve micro-disentanglement for representing each item. This method\citep{D-GCN} does not consider macro-separability between each pair of factors behind the item. Thus, DGCF\citep{DGCF} and IPGDN\citep{I-GCN} employ distance correlation and the Hilbert-Schmidt independence criterion, respectively, to encourage independence between pairs of factors. In SBR, Li et al.\citep{Disen-GNN} is the first work to consider different contributions of item factors to capture the user's intent and apply disentangled representation learning techniques to learn different factors embeddings of item in the current session, which infer the main intent of the user more easily. Capturing the main intents of users can effectively assist the model in inferring the reasons for transition relationships between items. Thus, we employ disentanglement techniques to represent each item as different factor embeddings to infer the key factor that the user cares about as the reason for the transition between items in the global context.
\subsection{Contrastive Learning}
Contrastive learning aims at embedding the extended versions of the same sample close to each other, and trying to push the embedded contents of different samples away \citep{survey}. It has become an indispensable component in improving model performance. Early contrastive learning works focus on word-embedding methods\citep{WE-C}\citep{WE-C1}. Then it is applied to computer vision\citep{SimCLR} and natural language processing \citep{CLEAR}\citep{Declutr}. In the field of graph,  DGI \citep{DGI} and DMI \citep{GMI} use comparative learning to enhance the robustness of node representation in graphs. In SBR, data sparsity is a problem that has been perplexing this field. Comparative learning can be seen as a method of data augmentation to enhance the robustness of representation. Some works \citep{DHCN} has combined comparative learning with session-based recommendation and achieve good results. To enhance the robustness of session representations, we adapt Contrastive Predictive Coding \citep{CPC} proposed infoNCE loss into our model.

%% file: 3-method.tex
\section{PRELIMINARIES}
In this section, we first introduce the problem setting for SBR, then present two types of graph models, i.e., position-aware global graph and session graph. We highlight the process of modeling fine-grained transition information over the global graph.

\begin{figure*}[t]
  \centering\includegraphics[scale=0.69]{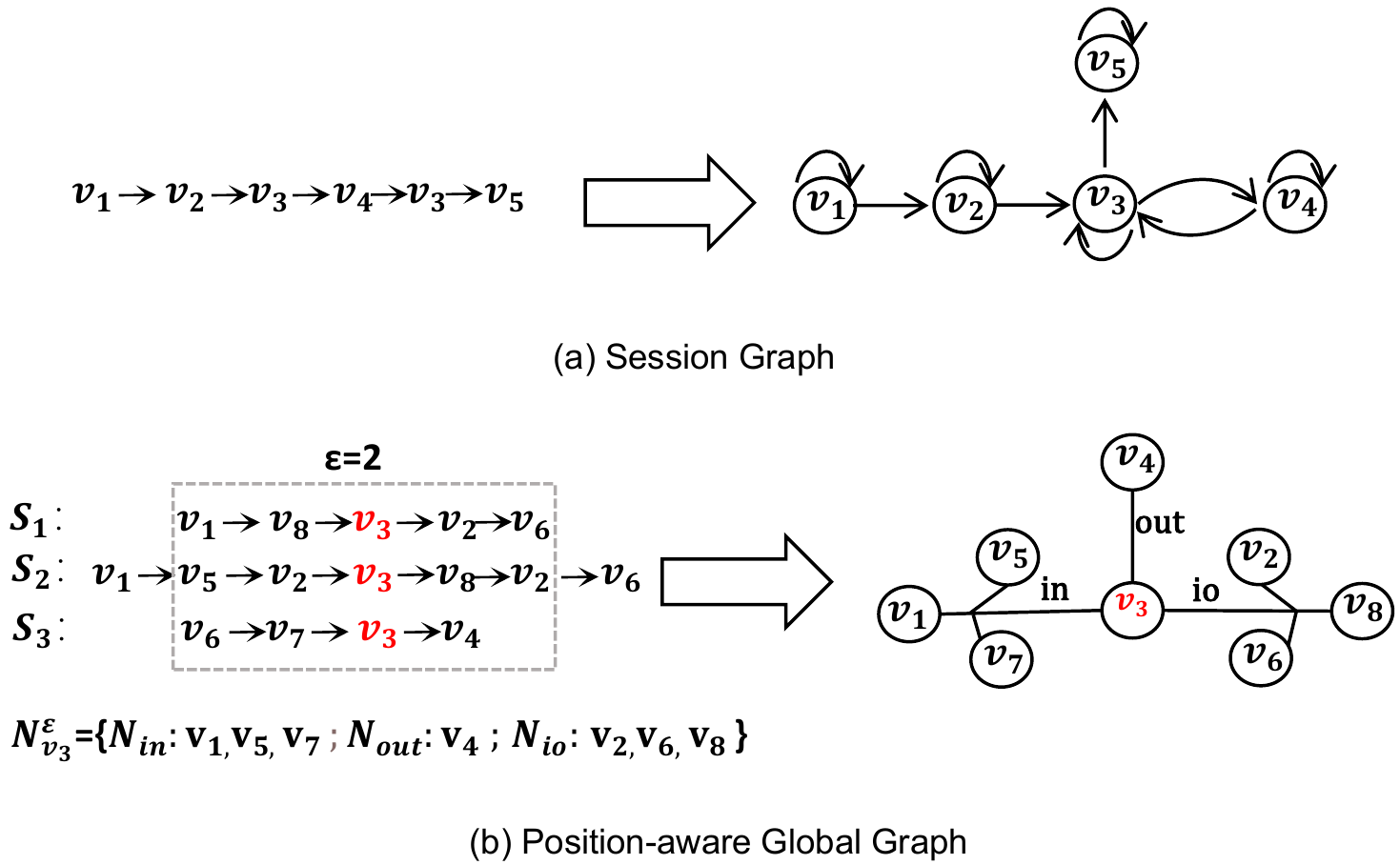}

  \centering
\caption{ \textrm{Illustrations of construction of session graph and position-aware global graph. }}
\label{graph}
\end{figure*}

\subsection{Problem Setting}
 Let $\mathcal{V} = \{v_1,... v_i,... v_N\}$ represent the set of all unique items involved in all sessions, and $N$ represents the total number of items. $s=[v_{s,1},v_{s,2},...,v_{s,n}]$ denotes an anonymous session in which items are ordered by timestamps and $v_{s,k}\in \mathcal{V}(1 \leq k \leq  n)$ denotes an interacted item by user within the session $s$. In our model, we embed each session s and item $v_{i}$ into the same space and let $\mathbf{s}$ and $\mathbf{v_{i}}$ represent them, respectively\footnote{ In the paper, we use bold uppercase letters, bold lowercase letters, and nonbold letters to denote matrices, vectors, and scalars, respectively. unless otherwise specified, all vectors are in the column form.}. 
 Given a session $s$, the session-based recommendation task is to recommend the next item $v_{s, n+1} \in \mathcal V$ that is most likely to be interacted with by the user of the current session $s$.
\subsection{Graph Model Construction}
\subsubsection{Construct Position-aware Global Graph}\label{global:graph}
In the learning process of a graph-based model, we extract information from the graph structure through node propagation. The information contained in the graph structure determines the upper limit of the performance of our model. Therefore, it is crucial to make the graph model contain as much information as possible. In session-based recommendation, some methods consider mining information from other sessions to construct a global graph to achieve better performance. Existing global graph construction methods, such as I3GN \citep{I3GN} and DGTN \citep{DGTN} select neighbors from recent sessions based on their similarities to each item in the target session. 
GCE-GNN \citep{GCE-GNN} traverses all session sequences to select neighbors that appear most frequently for each item. All these methods unify all the filtered neighbors into a single set without distinguishing the different types of neighbors. This will make it difficult for the model to capture item transition in the global context. Because there is no latent information to make the model obtain the common transition between neighbors, such as whether the transition is from \textit{A} to \textit{B} or \textit{B} to \textit{A} or both, in most cases. Thus, we propose a novel position-aware global graph, which utilizes the relative position information to distinguish the role of different neighbors, so as to model finer-grained transition information in the global context.

Let $\mathcal{G}_{v_i}^{g}=( \mathcal{V}_g,\mathcal{E}_g)$ be the position-aware global graph. First, we define $|\bm{\varepsilon}|$ as the scope of modeling of item transition, and traverse all sessions to find all the neighbor items, which the adjacent distance $d \le |\bm{\varepsilon}|$ to $v_i$, and all its neighbor items and itself to form the $\mathcal{V}_g$ node set in the global graph $\mathcal{G}_{v_i}^g$. These neighbors are represented as set $N_{v_i}^{\varepsilon}$. To modeling position information, there are three types of edges in edge set $\mathcal{E}_g$, namely in-edge, out-edge and in-out-edge. Based on this, we divide the items in the neighbor set $N_{v_i}^{\varepsilon}$ into $N_{in}, N_{out}$ and $N_{io}$, representing in-coming neighbor, out-coming neighbor and in-out-coming neighbor, respectively. Each type of neighbor is connected to the corresponding type of edge. In order to distinguish the importance of neighbors, we take the frequency of neighbor items over all the sessions as the weight of the corresponding edge. An example of building a position-aware global graph is shown in the Figure \ref{graph} (b).

\subsubsection{Construct Session Graph}
For each session sequence $s=[v_{s,1},v_{s,2},...,v_{s,n}]$, we construct a session graph $\mathcal{G}_{s}=(\mathcal{V}_{s},\mathcal{E}_{s})$ to model the pattern of neighboring items in the current session, where $\mathcal{V}_{s}$ and $\mathcal{E}_{s}$ are the node set and edge set, respectively. In our setting, each node represents an item $v_{s,i}\in \mathcal{V}_{s}$. $(v_{s,i},v_{s,j})\in \mathcal{E}_{s}$ indicates that there is an adjacent edge between nodes $v_{s,i}$ and $v_{s,j}$. Four types of edges are contained in our edge set, which are $e_{in}$,$e_{out}$,$e_{in-out}$ and $e_{self}$, respectively. $e_{in}$ indicates in-coming edge that there is a transition from $v_i$ to $v_j$, $e_{out}$ indicates out-coming edge, $e_{in-out}$ indicates in-out-coming edge that there are both transition from $v_i$ to $v_j$. $e_{self}$ denotes that there is a loop transition within the item itself.
These all types of edges can help model to capture the relationship between items at the session-level more easily. The examlple is shown in the Figure \ref{graph} (a).

\begin{figure*}[t]
  \centering\includegraphics[scale=0.38]{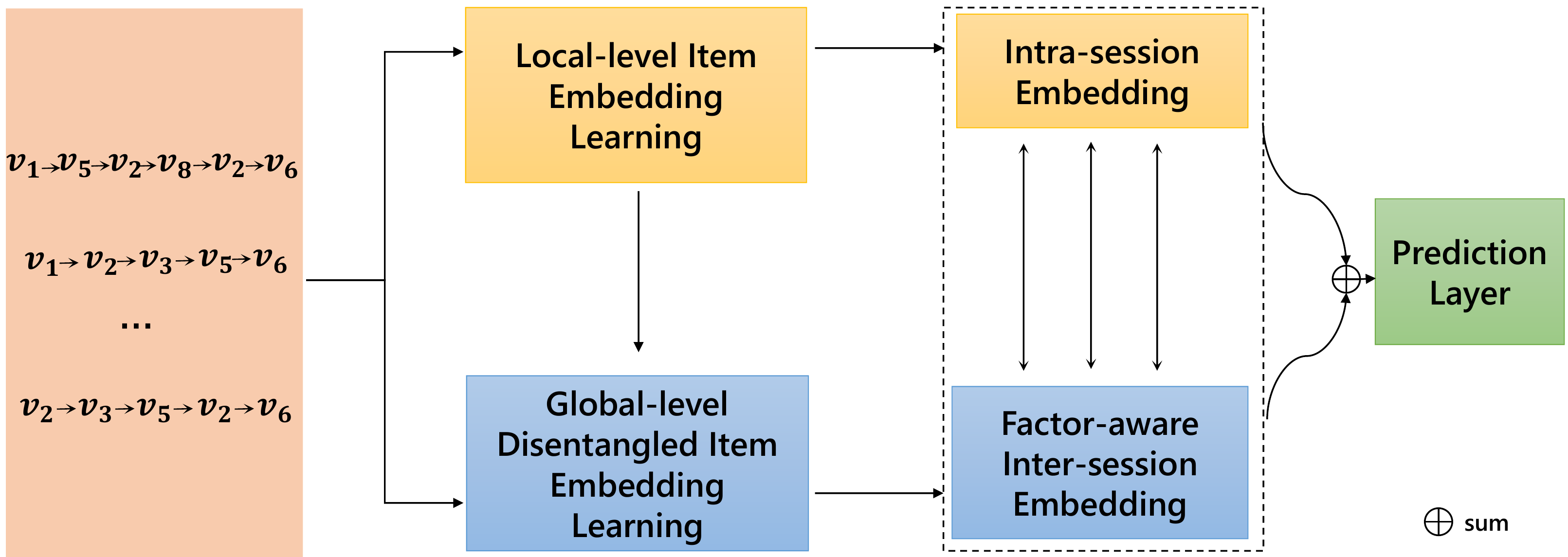}

  \centering
\caption{ \textrm{The overview of the proposed TIE-DGNN model. }}
\label{overview}
\end{figure*}

\section{THE PROPOSED METHOD}
In this section, we present our Transition Information Enhanced Disentangled Graph Neural Network model, termed TIE-DGNN, whose workflow is shown in Figure \ref{overview}. It is composed of four main components. 1) \textbf{Global-level Disentangled Item Embedding Learning. }In this module, we first initialize the position embeddings in the global context and slice item embeddings into multiple chunks with the assumption that each chunk represents a factor. Then, we employ global-level disentangling layers to separately learn different factor embeddings of items over the global graph and finally get the factor-aware global item embeddings. 2) \textbf{Local-level Item Embedding Learning. }In this module, we use an attention mechanism to learn the local-level item embeddings in the current session. 3) \textbf{Dual-channel Session Embedding Learning. }In this module, we take reversed position information into attention weights account to generate factor-aware inter-session embedding and intra-sesion embedding via the aggregation of factor-ware global item embeddings and local-level item embeddings, respectively. Then, we employ contrastive learning techniques to enhance the robustness of two types of session embeddings. 4) \textbf{Prediction Layer. }In this module, we linearly combine two types of session embeddings and match it with candidate items to calculate the probability of their being the next item. 
\subsection{Global-level Disentangled Item Embedding Learning}
In this subsection, we first introduce the initialization of item embeddings and position embeddings in the global context, and then show how to learn the factor-aware global item embeddings via global-level disentangling layers.
\subsubsection{Initialization}\label{init}
\textbf{Embedding Initialization.} The previous methods \citep{SR-GNN,GCE-GNN} of session-based recommendation represent an item as a holistic representation. However, the intents behind users' selection of items are diverse \citep{MCPRN,D-GCN}, which are determined by different potential factors behind the item.  Thus, we use disentangled embedding learning techniques to encode the different factors behind the item, so as to infer the key factors that the user most cares about for interpreting the reason of transition in the global context. Then, we take item $v_i$ as an example, showing how to generate initial disentangled item embeddings in the global context.

Disentangled representation learning aims to model the latent different factors behind the item. We assume that there are $K$ latent factors when given a single node $ i \in \mathcal{V}$ in the global graph. Thus, the embedding $\mathbf {v_i}\in \mathbb{R}^{d }$ is cast into K chunks, with each chunk representing a latent factor. The formula is as follows:
\begin{equation}
  \mathbf{c}_{i,k} = \frac{\sigma(\mathbf{W_{k}^{\top}} \cdot \mathbf{v_i})+\mathbf{b_{k}}}{{\|\sigma(\mathbf{W_{k}^{\top}} \cdot \mathbf{v_i})+\mathbf{b_{k}}\|}_2},\label{generate disentangled representation}
\end{equation}
where $\sigma$ is an activation function. $\mathbf{W_{k}} \in \mathbb{R}^{d \times \frac{d}{K} }$ is a weight matrix of the $k^{th}$ factors. $\mathbf{b_k}\in \mathbb{R}^{\frac{d}{K}}$ represents the bias term. $l_2$ normalization is adopted to avoid overfitting. Accordingly, the initial disentangled embedding for $v_i$ is $\mathbf{h}_{v_i^{(0)}}^g=\{\mathbf{c}_{i,1}^{(0)},...,\mathbf{c}_{i,k}^{(0)}\}\in\mathbb{R}^{\frac{d}{K}}$, where $\mathbf{c}_{i,k}$ represents embedding for the $k^{th}$ factor.

\textbf{Position Embedding Initialization.} As aforementioned, each item in the global graph has three types of neighbors, namely $N_{in},N_{out},N_{io}$, which represent in-coming neighbor, out-coming neighbor and in-out-coming neighbor, respectively. Different types of neighbors denote the different transition relationships for the current item. We consider that in the same transition relationship, the importance of different neighbors to the current item is differentiated, e.g., in the session $\{v_1 \rightarrow v_2 \rightarrow v_3 \rightarrow v_4\}$, $v_1,v_2,v_3$ are the in-coming neighbors for $v_4$. It is obvious that $v_3$ shows great influence on the current item $v_4$ and the impact of $v_1$ would be relatively small. The distance of neighbors from the current item can distinguish the importance of different neighbors. Thus, we propose a position-coding method by considering the distance to adaptively estimate the importance of each neighbor.

For neighbor sets $N_{in},N_{out}$, we construct two learnable position embedding  matrices $\mathbf{P}_{in},\mathbf{P}_{out}$. Take $\mathbf{P}_{in}=[\mathbf{p}_{1}^{in},...,\mathbf{p}_{\mu}^{in},...,\mathbf{p}_{\varepsilon}^{in}]\in \mathbb{R}^{d_p}$ as an example, $\mu$($ 1\leq \mu \leq \varepsilon$) denotes adjacent distance between items and $d_p$ represents the dimension of position embedding. In a dataset, the adjacent distance between neighboring items may not unified, e.g., in the sesion $\{v_1 \rightarrow v_2 \rightarrow v_3 \rightarrow v_4\}$, the distance of $v_3$ from $v_4$ is 1, however in session $\{v_3 \rightarrow v_9 \rightarrow v_4\}$ is 2. To better estimate the importance of each neighbor, we choose the most frequent distance of neighbors from the current item over all the sessions as $\mu$. For neighbor set $N_{io}$, the meaning of distance of in-coming edge and out-coming edge is different. Thus, we do not consider distance of in-out-coming neighbors from the current item. To distinguish other types of neighbors, we build a unified position vector $\mathbf{p}_{io}\in\mathbb{R}^{d_p} $ for all items in $N_{io}$. 

\begin{figure*}[t]
  \centering\includegraphics[scale=0.38 ]{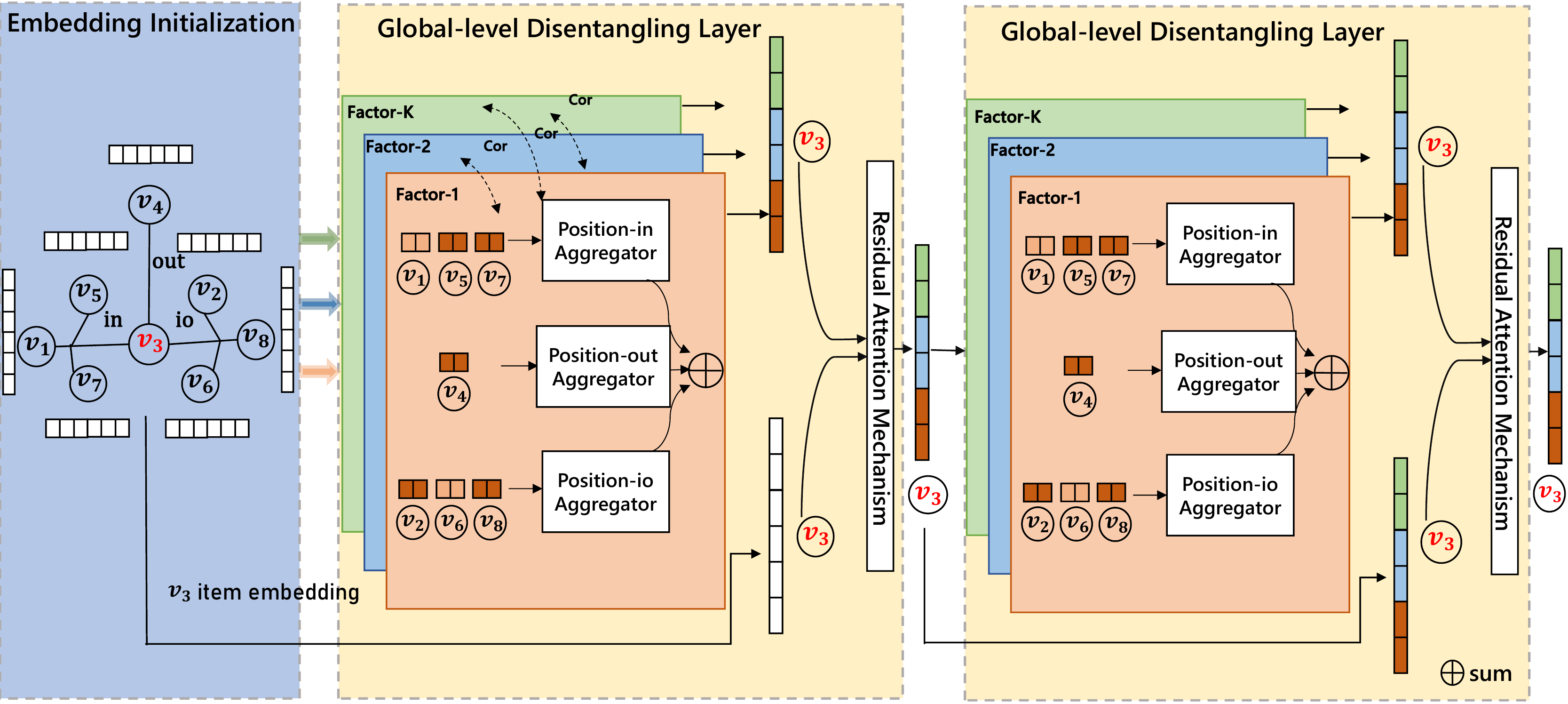}

  \centering
\caption{ \textrm{An example of two GLDL layers for our disentangled embedding learning method. Specifically, we first project each node (i.e., item) in the position-aware global graph into different embedding subspaces such that each subspace represents a latent factor. Then, in each subspace, we employ position-aware aggregators to separately aggregate neighbor information of different position types for learning factor embeddings of nodes. In the next stage, we take $v_3$ as an example. The different factor embeddings are concatenated to form a new item embedding. To prevent overfitting, we fused the new item embedding with its previous representation via the residual attention mechanism to generate the final item embedding. Finally, distance correlation is used to encourage independence between each pair of factors. Stack multiple GLDL layers can obtain high-order transition information between items over the graph. }}
\label{global}
\end{figure*} 


\subsubsection{Global-level Disentangling Layer}
In this subsection, we present details of global-level disentangling layer (\textbf{GLDL}). The example of structure of our model with two GLDL layers is shown in Figure \ref{global}. 

 \textbf{Position-aware Neighbor Information Propagation.} As mentioned above (section \ref{global:graph}), for each item, there is a neighbor set $N_{v}^{\varepsilon}=\{N_{in},N_{out},\\N_{io}\}$, which represent the in-coming neighbors, out-coming neighbors, and in-out-coming neighbors, respectively. Different types of neighbors play different roles in the learning process. We conduct information propagation for $N_{in},N_{out}$ and $N_{io}$, respectively. In the process of propagation, in order to distinguish the importance of different neighbors to the current items from the same type of neighbor, we employ an attention mechanism to achieve the goal. Moreover, we represent each item as different factor embeddings mentioned above (section \ref{init}). Thus, we separately propagate and update different factor embeddings of item. Let $L$ be the number of GLDL layers in our model and $K$ be the number of factors. We will use $l(1 \leq l \leq L)$\footnote{We omit the notation l for simplicity as the operation is the same for all the GLDL layers} and $k(1\leq  k\leq  K)$ to denote the $l$-th GLDL layer and $k$-th factor, respectively. We then present how to learn the global embedding of items.

To distinguish the importance of different items in the same neighbor type to the current item $v_i$, we employ attention mechanism and linearly combine neighbor information  according to attention score,
\begin{align}
\mathbf{h}_{N_{r_{i,j}^{g}}}^{k}=\sum_{v_{j}\in N_{{r_{i,j}^{g}}}}\theta_{i,j}\mathbf{c}_{j,k},\label{neighbor:1}
\end{align}
where $\mathbf{c}_{j,k}$ denotes the embedding of neighbor item $v_j$ for $k^{th}$ factor. $\theta_{i,j}$ represents the important weight of different neighbors in $N_{r_{i,j}^{g}}$ to current item $v_i$ where $r_{i,j}^{g}\in [in,out,io]$ represents the transition relationship between items $i,j$ in the global context\footnote{It should be noted that $N_{in}, N_{out}, N_{io} $ represents in-coming neighbors, out-coming neighbors and in-out-coming neighbors respectively, and we separately aggregate the neighbor information according to the transition relationship with attention weights.}. Intuitively, several factors can cause effects on the importance of a neighbor to the target item, such as the position information, the frequency infromation, and the matching degree with the session preference. We combine these three elements to estimate the important weight $\theta_{i,j}$ between neighbor $v_j\in N_{r_{i,j}^{g}}$ and the current item $v_i$, 
\begin{align}
  \theta _{i,j}=\mathbf{q}_{r_{i,j}^{g}}^{\top}LeakyRelu(\mathbf{W}_{r_{i,j}^{g}}[\mathbf{s}_k\odot \mathbf{c}_{j,k}\parallel w_{ij} \parallel \mathbf{p}_{\mu }^{r_{i,j}^{g}}]),\label{neighbor:2}
\end{align}  
where $\odot$ indicates element-wise multiplication operation, $\parallel$ indicates concatenation operation. LeakyRelu as the activation function. For different types of neighbor, we train two weight matrices $\mathbf{W}_{r_{i,j}^{g}} \in \mathbb{R}^{\frac{d}{K}+ d_p+1 \times \frac{d}{K}+d_p+1}$ and $\mathbf{q}_{r_{i,j}^{g}}\in \mathbb{R}^{\frac{d}{K}+ d_p+1}$. $\mathbf{p}_{\mu}^{r_{i,j}^{g}}\in \mathbb{R}^{d_p}$ is position vector to be learned where $\mu$ denotes distance between $v_j$ and $v_i$, and $d_p$ indicates the dimension of the vector. $w_{r_{i,j}^{g}}\in \mathbb{R}^{1}$ represents the weight between $v_i$ and $v_j$ which is determined by the number of $v_j$ occurrences. $\mathbf{s}_k$ can be seen as preference of the current session for the $k^{th}$ factor, which is as follows: 
\begin{align}
  \mathbf{s}_{k} =\frac{1}{|S|} \sum_{v_{i}\in S }\mathbf{h}_{v_i,k}^{s}.\label{neighbor:3}
\end{align}

$\mathbf{h}^{s}_{v_i,k}$ represents local-level item $v_i$ embedding for the $k^{th}$ factor obtained by the local-level representation learning layer. The details will be described in the subsequent section \ref{local_item}. 

We employ softmax function to normalize the coefficients from all neighbors in $N_{r_{i,j}^{g}}$ to $v_i$, which makes coefficients comparable across different neighbors: 
\begin{align}
  \theta _{i,j}= \frac{exp(\theta _{i,j})}{\sum _{v_{x}\in N_{in}} exp(\theta _{i,x})}.\label{neighbor:4}
\end{align}

The resulting attention score is capable of estimating the importance of each neighbor.

According to the above equations (\ref{neighbor:1}) (\ref{neighbor:2}) (\ref{neighbor:3}) (\ref{neighbor:4}), we aggregate the neighbor information in $N_{in}$, $N_{out}$, $N_{io}$ to compute $\mathbf{h}_{N_{in}}^{k}$, $\mathbf{h}_{N_{out}}^{k}$ and $\mathbf{h}_{N_{io}}^{k}$. Finally, we linearly combine the three kinds of neighbor information to get the final neighbor information representation:
\begin{align}
\mathbf{h}^{k}_{N_{v_i}^{\varepsilon }}=\mathbf{h}_{N_{in}}^{k}+\mathbf{h}_{N_{out}}^{k}+\mathbf{h}_{N_{io}}^{k}.
\end{align}

The current item pays different attention to different factor embeddings of neighbors. Therefore, for each factor, we get an aggregated neighbor information representation. $\mathbf{h}_{N_{v_i}^{\varepsilon }}=[\mathbf{h}^{1}_{N_{v_i}^{\varepsilon }},...,\mathbf{h}^{K}_{N_{v_i}^{\varepsilon }}]\in\mathbb{R}^{\frac{d}{K}}$ indicates the set of neighbor information embedding for all factors.

\textbf{Node update.} In this step, we will aggregate the neighbor information embedding and current item embedding $\tilde{\mathbf{h}}_{v_i}^{g}= \{\mathbf{c}_{i,1},...,\mathbf{c}_{i,k}\}\in\mathbb{R}^{\frac{d}{K}}$ factor-by-factor. The formula is as follows:
\begin{align}
  \mathbf{h}_{v_i,k}^{g}=relu(\mathbf{W}_{k_1}[\mathbf{c}_{i,k}\parallel \mathbf{h}^{k}_{N_{v_i}^{\varepsilon }}]),
\end{align}
where $\parallel$ indicates concatenate operation. We select relu as active function. $\mathbf{W}_{k_{1}}\in \mathbb{R}^{\frac{d}{K} \times \frac{2d}{K}}$ denotes the weight matrix to be learned. After current item is updated factor-by-factor, we obtain $\mathbf{h}_{v_i}^{g}=[\mathbf{h}_{v_i,1}^{g},...,\mathbf{h}_{v_i,K}^{g}] \in \mathbb{R}^{\frac{d}{K}}$. Then, let the learned embeddings of all factors concatenate into a holistic representation $\mathbf{h}_{v_i}^{g} \in \mathbb{R}^{d}$.

It is well known that in the information propagation of graph-based models, with the increase of the number of layers, there will be an indistinguishable problem between items, which is called over-smoothing. Inspired by \citep{Disen-GNN}, we employ \textbf{residual attention mechanism} to alleviate the negative effect of over-smoothing problem:
\begin{align}
  &\alpha =\mathbf{W}_f(\sigma(\mathbf{W}_p\mathbf{h}_{v_i}^{g}+\mathbf{W}_q\mathbf{h}_{v_i}^{g^{(l-1)}})),\\ 
  &\mathbf{h}_{v_i}^{g^{l}}=\alpha\mathbf{h}_{v_i}^{g}+(1-\alpha)\mathbf{h}_{v_i}^{g^{(l-1)}},
\end{align}
where $\mathbf{h}_{v_i}^{g^{(l-1)}}$ denotes the final output representation of $v_i$ in the $(l-1)$-th GLDL layer. $\mathbf{W}_p,\mathbf{W}_q \in \mathbb{R}^{d\times d}$ and $\mathbf{W}_f \in \mathbb{R}^{1\times d}$ are learnable parameters. $\sigma $ is the sigmoid activation function. $\alpha$ controls the amount of each part should be preserved.

\textbf{Distance Correlation.} As previously stated, in order to characterize the features of items, we cast item embedding into $K$ chunks, with each chunk representing a latent factor. We would like these factors to be independent to each other, which avoid the negative effects of information redundancy. In fact, the information redundancy always exists between these factors. We need take measures to alleviate the negative impact of it. Thus, we employ distance correlation as the regularizer in our model to further encourage the independence between each pair of factors. The formula is as follows:
\begin{align}
  \mathcal{L}_{cor}=\sum_{k=1}^{K} \sum_{k^{'}=k+1}^{K} cos(\mathbf{c}_{i,k},\mathbf{c}_{i,k^{'}} ),\label{cor}
\end{align}
where $cos$ represents the similar distance between two embeddings. $\mathbf{c}_{i,k},\mathbf{c}_{i,k^{'}}$ indicate a pair of factor embeddings of an arbitrary item $v_i$. For more details, please refer to \citep{MTCD}.

\subsection{Local-level Item Embedding Learning}\label{local_item}
To learn the local neighboring item transitions contained in the current session, inspired by \citep{GCE-GNN}, the edge set of the session graph contains four types of edges to model relationships between items, namely $e_{in},e_{out},e_{in-out},e_{self}$. Next, we will present how to learn local-level item embedding.\footnote{ In the session graph model, due to its sparse user behaviors and considering the complexity of the model, we do not employ disentangled embedding learning techniques and represent each item as a holistic embedding (i.e., $\mathbf{h}_{v_i} \in \mathbb{R}^{d}$).} It is well-known that the importance of different neighbors to the current item is different. Thus, we assign different attention weights to neighbors to distinguish their importance and linearly combine them:
\begin{align}
\mathbf{h}_{v_i}^{s}=\sum_{v_j\in N_{v_i}^{s}}\varphi_{i,j}\mathbf{h}_{v_j}.\label{local:1}
\end{align}

$\varphi_{i,j}$ represents the coefficient controlling the importance weight of neighbors, which is calculated by similarity of neighboring item:
\begin{align}
\varphi _{i,j}=LeakyRelu[\mathbf{W}_{r_{ij}^{s}}(\mathbf{h}_{v_i}\odot \mathbf{h}_{v_j} ) ],\label{local:2}
\end{align}
where $\odot$ indicates the element-wise product. We choose LeakyRelu as activation function. It should be noted that there are four types of edge relationship in the session graph. Thus, the four weight matrices need to be learned to correspond to the four edge relationships, namely $\mathbf{W}_{in},\mathbf{W}_{out},\mathbf{W}_{in-out},\\\mathbf{W}_{self}$. $\mathbf{W}_{r_{i,j}^s} \in \mathbb{R}^{1 \times d}$ denotes weight matrix to be learned where $r_{i,j}^{s}\in [in,out,in-out,self] $ represents the relationship between $v_i$ and $v_j$.  

Then we apply the softmax function to normalize the importance weights across all neighbors (including itself) to the current item:
\begin{align}
\varphi _{i,j}=\frac{exp(\varphi _{i,j})}{ {\textstyle \sum_{v_{x}\in N_{v_i}^{s}}exp(\varphi _{i,x})^{}} }.\label{local:3}
\end{align}

According to Eq (\ref{local:1})(\ref{local:2})(\ref{local:3}), we obtain local-level item embedding $\mathbf{h}_{v_i}^{s} \in \mathbb{R}^{d}$ which is aggregated by the features of neighbors and item itself in the current session. 

\textbf{Remark.} It should be noted that Eq (\ref{neighbor:3}) in the global-level disentanglement layers need obtain the trained embedding of items at the local-level to generate the average feature representation of the current session for different factors. Therefore, we allow the trained embeddings of each item $\mathbf{h}_{v_i}^{s}\in \mathbb{R}^{d}$ in the current session to be cast into $K$ embedding subspaces that train the factor embeddings in the global context, obtaining a set $\mathbf{h}_{v_i}^{s}=[\mathbf{h}_{v_i,1}^{s},...,\mathbf{h}_{v_i,K}^{s}]\in \mathbb{R}^{\frac{d}{K}}$. 

\subsection{Dual-channel Session Embedding Learning}
For each item in an arbitrary session sequence, we learn two types of embeddings: global-level (inter-session) disentangled item embedding $\mathbf{h}_ {v_i} ^ {g} $ and local-level (intra-session) item embedding $\mathbf{h}_ {v_i}^{s}$. The previous methods \citep {GCE-GNN} \citep {DGTN} fuse two different levels of item embedding before learning session embeddings, however, it may introduce more noise into the embedding. Thus, we decided to separately learn session embedding at two different views (i.e, inter-session and intra-session embeddings). Next, we will present how to generate two different views of session embedding. 

\textbf{Factor-aware Inter-session Embedding Learning.}  
In this subsection, we will show how to learn inter-session embedding. It is well-known that session embedding directly aggregated by the item which is represented as a holistic embedding cannot well model the diverse intents of a user in the current session\citep{Disen-GNN}. Based on factor-aware item embeddings obtained by global-level disentangling layers, we can assign attention coefficients to different factors of each item in the process of generating factor-aware session embeddings. By doing so, we can estimate the diverse intents of a user on different factors. This is expected to help us better capture the main factors that users care about and infer the key reason for the main transition relationships in the current session.

In addition, how to assign the weight coefficient is also particularly important for each item in the current session. Most previous methods focus on the importance of the last item in the session, which indirectly affects the contribution of other items to the current item. We need to measure the contribution of each item in the session more comprehensively. Intuitively, the importance of each item decreases from the back to the front of the session. Thus, we integrate reversed position embedding into the process of assigning weight coefficient to each item.

Let $\mathbf{h}_{v_i}^{g}=[\mathbf{h}_{v_i,1}^{g},...,\mathbf{h}_{v_i,K}^{g}] $, obtained by global-level disentangling layers, be the input to represent each item in the current session. We also use a learnable position embedding matrix $\mathbf{P}_g=[\mathbf{p}_1^g,\mathbf{p}_2^g,...,\mathbf{p}_{\iota }^g]\in \mathbb{R}^{\frac{d}{K}} $ to model the reversed position information of item, where $\iota$ represents the length of session sequence. The $k^{th}$ factor embedding of the $t^{th}$ item in the session after fusing the position information is as follows:
\begin{align}
\mathbf{h}_{v_i,k}^{g^{'}}=tanh(\mathbf{W}_{k_2}[\mathbf{h}_{v_i,k}^{g}\parallel \mathbf{p}_{\iota-i+1}^g]+\mathbf{b}_{k_1} ),\label{inter:0}
\end{align}
where $\mathbf{W}_{k_2} \in \mathbb{R}^{\frac{d}{K} \times \frac{2d}{K}}$, $ \mathbf{b}_{k_1} \in \mathbb{R}^{\frac{d}{K}}$ are the trainable parameters. Then we calculate the representation of average feature of session,
\begin{align}
\mathbf{s}_{f,k}=\frac {1}{\iota} \sum_{i=1}^{\iota} \mathbf{h}_{v_i,k}^{g}.\label{inter:1}
\end{align}

The soft-mechanism is then employed to calculate the weight coefficient of each item,
\begin{align}
\gamma _{i} = \mathbf{q}_{k}^{\top}\sigma (\mathbf{W}_{k_3}\mathbf{h}_{v_i,k}^{g^{'}}+\mathbf{W}_{k_4}\mathbf{s}_{f,k}+\mathbf{b}_{k_2}),\label{inter:2}
\end{align}
where $\mathbf{W}_{k_3},\mathbf{W}_{k_4} \in \mathbb{R}^{\frac{d}{K}\times \frac{d}{K} }$ and $\mathbf{q}_{k},\mathbf{b}_{k_2} \in \mathbb{R}^{\frac{d}{K}}$ are learnable parameters. Finally, we linearly combine the item embeddings $\mathbf{h}_{v_i,k}^{g}$ to obtain the inter-session embedding on the $k^{th}$ factor:
\begin{align}
\mathbf{s}_{g,k}=\sum_{i=1}^{\iota } \gamma _{i}\mathbf{h}_{v_i,k}^{g}.\label{inter:3}
\end{align}

Following the above steps, the inter-session embedding can be obtained by factor-wisely aggregating all the item embeddings in the current session with weight coefficients.  $\mathbf{s}_g=[\mathbf{s}_{g,1},\mathbf{s}_{g,2},...,\mathbf{s}_{g,K}]\in \mathbb{R}^{\frac{d}{K}}$ represents the final factor-aware inter-session embedding.

\textbf{Intra-session Embedding Learning.}    
In this subsection, we use item embedding obtained by the Local-level Item Embedding Learning to learn the intra-session embedding. We adopt the same strategy as generating inter-session embedding, which is combined with position information to more comprehensively model the contribution of each item in the session. The difference is that we do not need to aggregate item embedding factor-wisely because we do not investigate latent factors at local-level item embedding learning due to the limited user behaviors.
Therefore, we only need to construct a new learnable position embedding matrix and adjust some learnable parameters in Eq (\ref{inter:0})(\ref{inter:1})(\ref{inter:2})(\ref{inter:3}), then generate the final intra-session embedding $\mathbf {s}_ {l}\in \mathbb{R}^{d}$ according to these formulas. Specifically, we construct a learnable position embedding matrix $\mathbf{P}_{l}=[\mathbf{p}_1^{l},\mathbf{p}_2^{l},...,\mathbf{p}_{\iota }^{l}] \in \mathbb{R}^{d}$ where $\iota$ denotes the length of the current session. For Eq (\ref{inter:0}), we apply $\mathbf{p}_{\iota-i+1}^l$ to be the position vector and $\mathbf{W}_{k_5}\in \mathbb{R}^{d \times 2d}$, $\mathbf{b}_{k_3}\in \mathbb{R}^{d}$ to be the learnable parameters. For Eq (\ref{inter:2}), we apply $\mathbf{W}_{k_6},\mathbf{W}_{k_7}\in \mathbb{R}^{d \times d}$ and $\mathbf{q}_{k_1},\mathbf{b}_{k_4} \in \mathbb{R}^{d} $ to be the learnable parameters.

\textbf{Contrastive learning.}  
Contrastive learning has been widely used in SBR, which can be used as an auxiliary task to enhance the performance of model. It is expected to better characterize different aspects of sessions by contrasting two groups of session embeddings learned via two views (inter-session and intra-session). Thus, we next present how to learn contrast objective to enhance the performance in characterizing session feature.

We first generate sample pairs from the ground truth (positive) and the corrupted samples obtained by corrupting positive samples with row-wise and column-wise shuffling. Then, we employ InfoNCE \citep{CPC} with a standard binary cross-entropy loss as our learning objective and the formula defined as follows:
\begin{align}
  \mathcal{L}_{con}=-log\sigma(\mathbf{s}_{g}\mathbf{s}_{l}^{\top})-log\sigma(1-(\mathbf{\tilde{s}}_{g}\mathbf{s} _{l}^{\top})),\label{con}
\end{align}
where $\mathbf{\tilde{s}}_{g}$ (or $\mathbf{\tilde{s}}_{l}$) represents the corrupted samples. By doing so, the session embedding can leverage another view of session information to refine itself and enhance robustness.

\subsection{Prediction Layer}
After obtaining inter-session embedding $\mathbf {s}_g $ and intra-session embedding $\mathbf{s}_l $, we linearly combine them into the final session representation:
\begin{align}
\mathbf{S}= \mathbf{s}_g+\mathbf{s}_l.
\end{align}

Based on the current session representation $\mathbf{S}$ and the embedding of candidate item $\mathbf{v_i}\in \mathcal{V} $, the probability of the next click is obtained by the dot product of their embeddings and applying the softmax function:
\begin{align}
\hat{\mathbf{y}}_i =Softmax(\mathbf{S}^{\top}\mathbf{v}_i). 
\end{align}

We employ a cross-entropy loss function as the learning objective, which is defined as:
\begin{align}
\mathcal{L}_c=-\sum_{i=1}^{n}\mathbf{y}_ilog(\hat{\mathbf{y}}_i)+(1-\mathbf{y}_i)log(1-\hat{\mathbf{y}}_i),  
\end{align}
where $\mathbf{y}$ is the one-hot encoding vector of the ground truth item. Then, we unify disentangled loss, contrastive loss and cross-entropy loss into the learning objective. The final loss function of our model is defined as follows:
\begin{align}
  \mathcal{L}=\mathcal{L}_c+\beta\mathcal{L}_{cor}+ \lambda\mathcal{L}_{con},\label{final}
\end{align}
where $\mathcal{L}_{cor},\mathcal{L}_{con}$ denotes disentangled loss and contrastive loss respectively which are defined in Eq. \ref{cor} and Eq. \ref{con}. 
$\beta$ and $\lambda$ controls the magnitude of the disentangled learning task (i.e., distance correlation loss) and contrastive learning task (i.e., InfoNCE loss), respectively.

%% file: 4-experiment.tex
\section{Experiments}
To evaluate the effectiveness of our proposed model, we conduct extensive experiments on three publicly accessible datasets by answering the following three key research questions:
\begin{itemize}[leftmargin=*]
  \item{\textbf{RQ1:}} Does the proposed method outperform state-of-the-art session-based baselines in publicly accessible datasets?

  \item{\textbf{RQ2:}} Does position-aware global graph and position embedding matrices in global-level encoder positively affect our proposed model on session-based recommendation?

  \item{\textbf{RQ3:}} How does the key parameters affect the performance of TIE-DGNN, including the number of factors for disentangled representation and the regularization coefficients for loss functions?

\end{itemize}
\begin{table}
	\centering
	\caption{Statistical results of datasets}
	\scalebox{0.80}{
  \resizebox{\columnwidth}{!}{
	\begin{tabular}{c c c c}
    \hline	
    \toprule
		Statistics & {\it Tmall} & {\it  Last.fm} & {\it Nowplaying} \\ \midrule
		\# training sessions & 351,268 & 2,837,644 & 825,304  \\
		\# test sessions & 25,898 & 672,519 & 89,824 \\
		\# items & 40,728 & 38,615 & 60,417  \\
		Avg.length & 6.69 &  11.88  & 7.42   \\
		\bottomrule
    \hline
	\end{tabular}
	}}
	\label{tab:dataset-statistics}
\end{table}
\subsection{Experimental Configurations}
\textbf{Datesets and Preprocessing. }We evaluate the performance of our proposed model on three real-world benchmark datasets:
\begin{itemize}[leftmargin=*]
  \item{\textit {Tmall} \footnote{\url{https://tianchi.aliyun.com/dataset/dataDetail?dataId=42 }}} dataset comes from IJCAI-15 competition, which contains anonymous user’s shopping logs on the Tmall online shopping platform. Due to the large size of Tmall, following \citep{GCE-GNN}, only the first 120000 of the sessions are used for experiments. In these sessions, we set the last 100 seconds as the test data and the remaining historical sessions as training data.
  \item {\textit {Last.fm} \footnote{\url{http://ocelma.net/MusicRecommendationDataset/lastfm-1K.html}}} dataset is widely used in the music recommendation task, which is released by \citep{lastfm}. We focus on the recommendation task for music artists. Following \citep{lastfm1} \citep{lastfm2}, the top 40,000 most popular artists are preserved, and the splitting interval is set to 8 hours. The most recent 20\% of the sessions are used as the test data and the remaining historical sessions as training data.
  \item {\textit {Nowplaying} \footnote{\url{http://dbis-nowplaying.uibk.ac.at/\#nowplaying}}} dataset describes the behaviors of user listening music extracted from Twitter, which comes from \citep{NOW}. Following \citep{GCE-GNN}, the sessions of the last two months are set to the test data and the remaining historical data for training.
\end{itemize}

For fair comparison, we conduct steps of preprocessing over the three datasets. Following \citep{SR-GNN} \citep{GCE-GNN}, we filter out all sessions which length is 1 and occurrences of items less than 5 times. In addition, we generate sequences and corresponding labels by using splitting method to augment data. To be specific, for a session sequence $S=[v_{s,1},v_{s,2},...,v_{s,n}]$, we generate a series of sequences and labels $([v_{s,1}],v_{s,2}),\\([v_{s,1},v_{s,2}],v_{s,3}), ...,([v_{s,1},...,v_{s,n-1}],v_{s,m})$. The statistics of the datasets are summarized in Table \ref{tab:dataset-statistics}.

\textbf{Evaluation Metrics. }To evaluate the recommendation results, we choose the most commonly used \textbf{P@20} (Precision) and \textbf{MRR@20} (Mean Reciprocal Rank) as metrics accrodding to the previous works\citep{GCE-GNN}\citep{SR-GNN}.

\textbf{Baseline Algorithms. }
We compare our method with the representative methods in SBR. The following eight baseline models are evaluated.
\begin{itemize}[leftmargin=*]
  \item{\textbf{FPMC\citep{FPMC} }}is a sequential method based on matrix factorization and Markov chain. To compare it in SBR, we ignore the user latent representations when conducting recommendation task. 

  \item{\textbf{GRU4Rec\citep{SR-RNN} }}is RNN-based method that utilizes GRU units to capture the sequential behaviors between items in session sequence.  
  
  \item{\textbf{NARM\citep{NARM} }}applies attention mechanism into hierarchical RNN to model the main purpose of user and combines it with sequential behavior to generate the representation for SBR. 
  
  \item{\textbf{STAMP\citep{STAMP} }}employs attention layers to replace all RNN encoders in the previous work, which is capable of capturing the users' current interests relied on last item and combines it with long-term interests to enhance the performance.  
  
  \item{\textbf{SR-GNN\citep{SR-GNN} }}utilizes the gated graph convolutional layer to obtain item embedding. Then, generating the session representation for recommendation by using an attention net which capture the global preference and current interests of this session. 

  \item{\textbf{GC-SAN\citep{GC-SAN} }}first combines graph neural network and multi-layer self-attention network to enhance the recommendation performance by modeling local neighboring item transitions and contextualized non-local representations.

  \item{\textbf{GCE-GNN\citep{GCE-GNN} }}uses two levels of graph models to capture item transition relationships from local and global contexts and takes reversed position information into account to generate session representation for SBR.

  \item{\textbf{Disen-GNN\citep{Disen-GNN} }}combines disentangled representation lear-\ ning techniques with gated graph convolutional layers to model latent factors behind the item to estimate the diverse intents of the user.
\end{itemize}

\textbf{Hyperparameter Setup. }For a fair comparison, we tune the baselines for best performance on three datasets according to the data preprocessing methods and the parameter settings provided in their papers. The Back-Propagation Through Time (BPTT) algorithm \citep{BPTT} is used for training our model. We adopt the dropout strategy in dual-channel session embedding learning to prevent overfitting.
Following previous methods \citep{GCE-GNN}\citep{SR-GNN}, we set mini-batch to 100 and L2 penalty to $10^{-5}$. All parameters are initialized using a Gaussian distribution with a mean of 0 and a standard deviation of 0.1. We select Adam with an initial learning rate of 0.001 that will decay by 0.1 after every 3 epoch to optimize parameters. Besides, we randomly select a 10\% subset of the training set as the validation set. Moreover, the scope of modeling of item transition $\varepsilon $ and number of neighbors in the global context is set to 3 and 12 respectively as previous methods \citep{GCE-GNN}. Some key parameters, which have a great impact on the model (i.e., the number of factors, the number of GLDL layers, etc.), need to be tuned respectively in each dataset. The details are shown in Table \ref{tab:Parameter}. 

\begin{table}
	\centering
	\caption{Parameter settings for datasets}
	\begin{threeparttable}
    \scalebox{0.80}{
    \resizebox{\columnwidth}{!}{
	\begin{tabular}{c c c c}
    \hline	
    \toprule
		Statistics & { \it Tmall} & { \it Last.fm } & { \it Nowplaying} \\ \midrule
		 embedding size (d) & 275 & 128 & 105  \\
		 \# factors (K) & 5 & 4 & 7  \\
    \# GLDL layers    & 2 & 2 & 2\\
     $\beta$ & 5 &  4  & 5 \\
     $\lambda$  & 0.005 &  0.02  & 0.005 \\
		\bottomrule
    \hline
	\end{tabular}
	} }
	\end{threeparttable}

	\label{tab:Parameter}
\end{table}

\begin{table}
	\centering
	\caption{The comparison of TIE-DGNN with baselines over the three datasets}
	\scalebox{0.9}{
  \resizebox{\columnwidth}{!}{
	\begin{tabular}{c|cc|cc|cc}
    \hline
	{\multirow{2}{*}{Model}} & \multicolumn{2}{c|}{Tmall} & \multicolumn{2}{c|}{Last.fm} & \multicolumn{2}{c}{Nowplaying} \\\cline{2-7} 
       &P@20 & MRR@20  & P@20 & MRR@20  & P@20 & MRR@20 \\ \hline
    FPMC  & 9.15 & 3.31 & 12.86 & 3.78 &7.36&2.82 \\
    GRU4REC &10.93&5.89&17.61&6.62 &7.92&4.48\\
    NARM &23.30&10.70 &21.83&7.59 &18.59&6.93\\
    STAMP  & 26.47 &13.36 &21.76&7.66&17.66&6.88 \\
    SR-GNN &27.57&13.72&22.33&8.23 &18.87&7.47 \\
    GC-SAN & 21.80&10.17&22.64&8.42 & 18.85& 7.43\\
    Disen-GNN  &31.56 & 15.31  &22.92 &{\bf8.75} & 22.22 &8.22 \\
    GCE-GNN & {\bf 35.09}&{\bf15.80} &{\bf24.26} &8.66 & {\bf22.47}&{\bf 8.40} \\ 
    \hline
    TIE-DGNN & {\bf 39.01}& {\bf 17.75} & {\bf 25.25}& {\bf 8.83}& {\bf 23.35}& {\bf 8.53}\\
    \hline
    \% Gain  &{\bf 11.2\%}  	&{\bf 12.3\%} &{\bf 4.0\%} &{\bf 0.9\%}  &{\bf 3.9\%} &{\bf 1.5\%} \\
    \hline
  \end{tabular}
	}}
	\label{tab:result-baseline-algorithms}
\end{table}

\subsection{Performance Comparison (RQ1)}
The performance comparison of our proposed model over the state-of-the-art baselines is shown in Table \ref{tab:result-baseline-algorithms}, where the best and second-best performance of each column are highlighted in boldface. The gains are calculated by using the difference between the performance of TIE-DGNN and the best baseline to divide the performance of the latter. It can be seen that our model has achieved the best results over the three datasets in terms of the two metrics consistently, especially improving a large margin on the Tmall dataset. The results ascertain the superiority of our proposed model.

As a traditional method in session-based recommendation, FPMC can only model the sequential behavior between two pairwise items for recommendation tasks. Thus, their performance is worse than that of neural network-based models (i.e., GRU4REC,NARM,STAMP).

Among all the neural network-based models, RNN is applied to build \\GRU4REC and NARM. While the performance of NARM is better than that of GRU4REC. This is because NARM employs an attention mechanism to capture the users’ current interests, which relies on the last item and combines it with sequential behavior obtained by RNN to conduct recommendation. While GRU4REC employs a simple RNN-based structure to consider sequential behavior between items and cannot model the shift of user preference. STAMP utilizes a complete attention mechanism that distinguishes the importance of different items and considers the last item of a session to model the short-term interest. Except for the improvement over NARM in Tmall dataset, the performance in other datasets is comparable.

Among all the baseline methods, the  graph-based methods  outperform other methods over the three datasets \ (besides GC-SAN in the Tmall dataset), which indicates that graph modeling is more suitable than sequential modeling (MC, RNN) or attention modeling for session-based recommendation. This is because the graph structure can iteratively aggregate the neighbor information from the graph to distill high-order transitions between items, which obtain better item representations. SR-GNN is the first work to employ gated graph neural network for session-based recommendation and use self-attention mechanism on the last item to generate session embedding. Based on SR-GNN, GC-SAN designs a multi-layer self-attention network to obtain contextualized non-local representations and combine them with the local item embeddings learned by GGNN for the recommendation task. Compared with SR-GNN, the performance of GC-SAN is comparable across the Last.fm and Nowplaying datasets. While in the Tmall dataset, it underperforms SR-GNN by a large margin, even lower than the neural network-based models (NARM, STAMP). We think that the multi-layer self-attention network may not be applicable to the data pattern in the Tmall dataset, which does not mean that graph neural networks are inferior to neural network-based models for SBR. Considering that the user's intents may be diverse in the session sequence, Disen-GNN disentangles the item embedding into independent factors that represent user preferences and separately updates the embedding of each factor via promotion over the session graph so as to better infer the user's specific intent. Attributed to the advanced representation learning techniques, Disen-GNN obtains better results than SR-GNN and GC-SAN across the three datasets. GCE-GNN is the best performing method that integrates information from global context (i.e., all the sessions) and current interests (i.e., current session) to learn item embeddings and also uses position information to generate session representation for SBR.

In our model, we highlight the importance of transitions between neighboring items from global and local contexts. Especially in the global context compared with GCE-GNN, we consider the finer-grained transition information, including its modeling, propagation, and the key reason for it. Specifically, we integrate relative position information into the global graph (global context) for each item to distinguish the different roles of neighbors. Then, we disentangle the item embedding into independent factors (represent user preferences) and separately update the embedding of each factor via promotion over the position-aware global graph. Combining them with local-level item embeddings to generate factor-aware inter-session embedding and intra-session embedding, respectively, and then using contrastive learning techniques to enhance the robustness of the two types of session embeddings. With those specially designed components, our model outperforms the best compared method (GCE-GNN) across the three datasets. In these three datasets, the improvement of our model in the Nowplaying and Last.fm datasets is relatively lower than in the Tmall dataset. We think it may be related to the average length of the session. The longer the session length, the richer the information contained in the session. In other words, even without considering finer-grained transition information, other models can still achieve better results. It also indicates that our model has greater potential in datasets with more sparse data through fine-grained modeling transition information, which is more in line with session-based recommendations.

\begin{table}
	\centering
	\caption{The effectiveness of global context}
	\scalebox{0.95}{
  \resizebox{\columnwidth}{!}{
	\begin{tabular}{c|cc|cc|cc}
    \hline
	{\multirow{2}{*}{Model}} & \multicolumn{2}{c|}{Tmall} & \multicolumn{2}{c|}{Last.fm} & \multicolumn{2}{c}{Nowplaying} \\\cline{2-7} 
       &P@20 & MRR@20  & P@20 & MRR@20  & P@20 & MRR@20 \\ \hline
       TIE-DGNN-w/o-PEM  &  37.52 & 16.90 & 24.97 & 8.65 & 21.97&8.45\\
       TIE-DGNN-w/o-PGG  & 38.10 & 16.77 & 24.12 & 8.30 & 23.03&7.66 \\
    TIE-DGNN & {\bf 39.01}& {\bf 17.75} & {\bf 25.25}& {\bf 8.83}& {\bf 23.35}& {\bf 8.53}\\
    \hline
  \end{tabular}
	}}
	\label{tab:result-position-aware global graph}
\end{table}

\subsection{ Effectiveness of global context (RQ2)}
\textbf{ Effectiveness of position-aware global graph. }
One of the main contributions of our model is designing the position-aware global graph at item-level to model the fine-grained transition relationships of neighboring items in the global context (i.e., all the sessions). Compared to the global graph constructed by GCE-GNN, we utilize position information to distinguish different types of neighbors (i.e., in-coming neighbors, out-coming neighbors and in-out-coming neighbors) and aggregate the neighbor information by position-wisely. To investigate the effectiveness of the position-aware global graph, we compared it with TIE-DGNN-w/o-PGG, which replaces the position-aware global graph with the global graph constructed by GCE-GNN. 

Table \ref{tab:result-position-aware global graph} shows the results of TIE-DGNN and the variant (TIE-DGNN-w/o-PGG). Our proposed model obtains better performance across three datasets. It indicates that by using position information to distinguish neighbors in the global context, TIE-DGNN learns the importance of different types of neighbors to better model the neighboring item transitions and thus performs better in the final recommendation task.

\textbf{ Effectiveness of position embedding matrices. }In the process of neighbor information aggregation, we train different position embedding matrices to estimate the importance of different neighbors under the same transition relationships and facilitate our model to capture neighboring item transitions in the global context. To further investigate the effectiveness of the position embedding matrices in the global context, we compare our proposed model with the variant (TIE-DGNN-w/o-PEM) that does not employ position embedding matrices in the aggregation of neighbor information. Table \ref{tab:result-position-aware global graph} displays the result. It is obvious that our proposed model obtains better performance across the three datasets, which demonstrates the positive effects of position embedding matrices on modeling neighboring item transitions in the global context.

\begin{figure*}[t]
	\centering
	\subfloat[P@20 on Last.fm]{\includegraphics[width=0.4\linewidth]{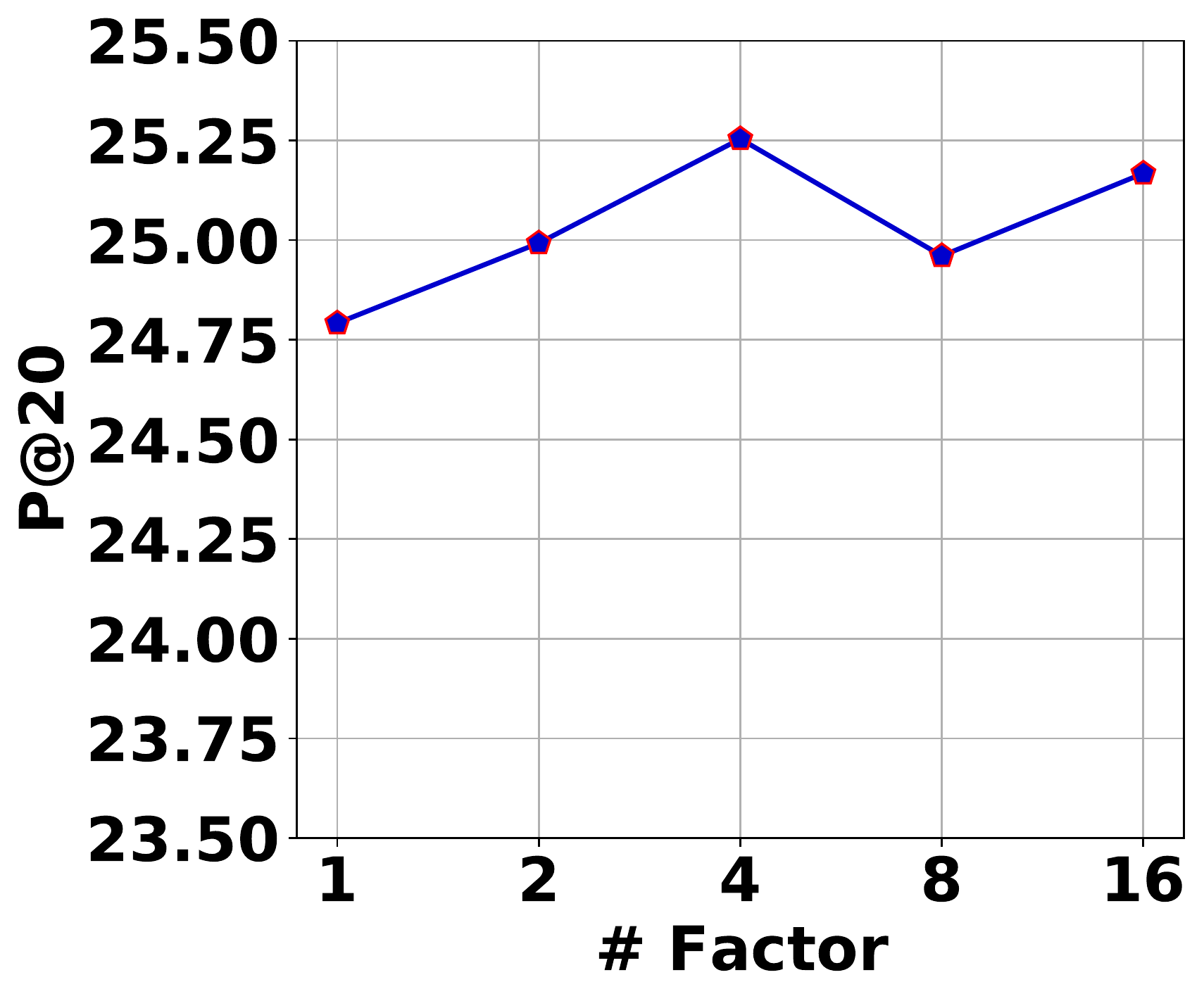}}
	\hspace{25pt}
  \subfloat[MRR@20 on Last.fm]{\includegraphics[width=0.37\linewidth]{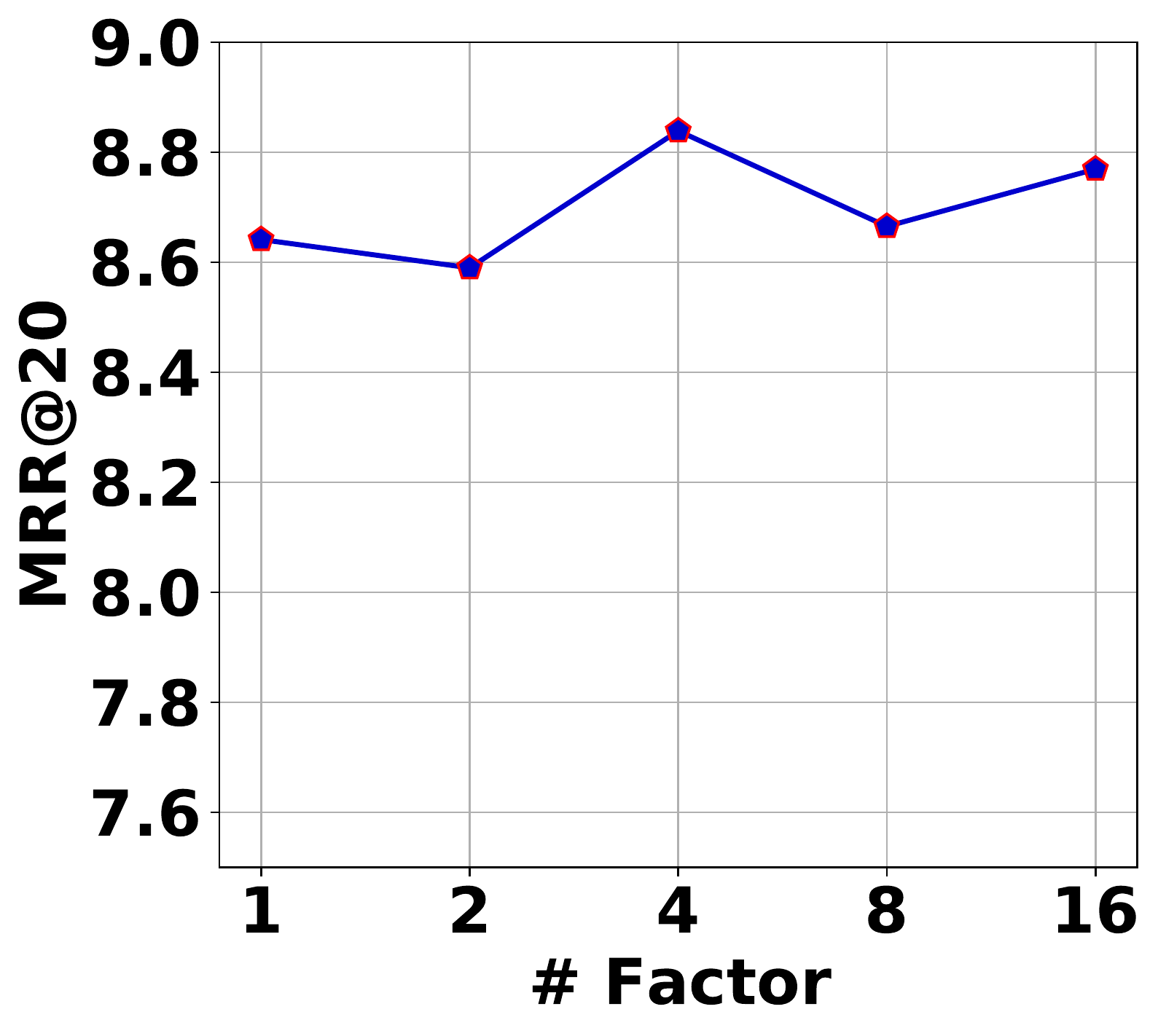}}

	\subfloat[P@20 on Nowplaying]{\includegraphics[width=0.4\linewidth]{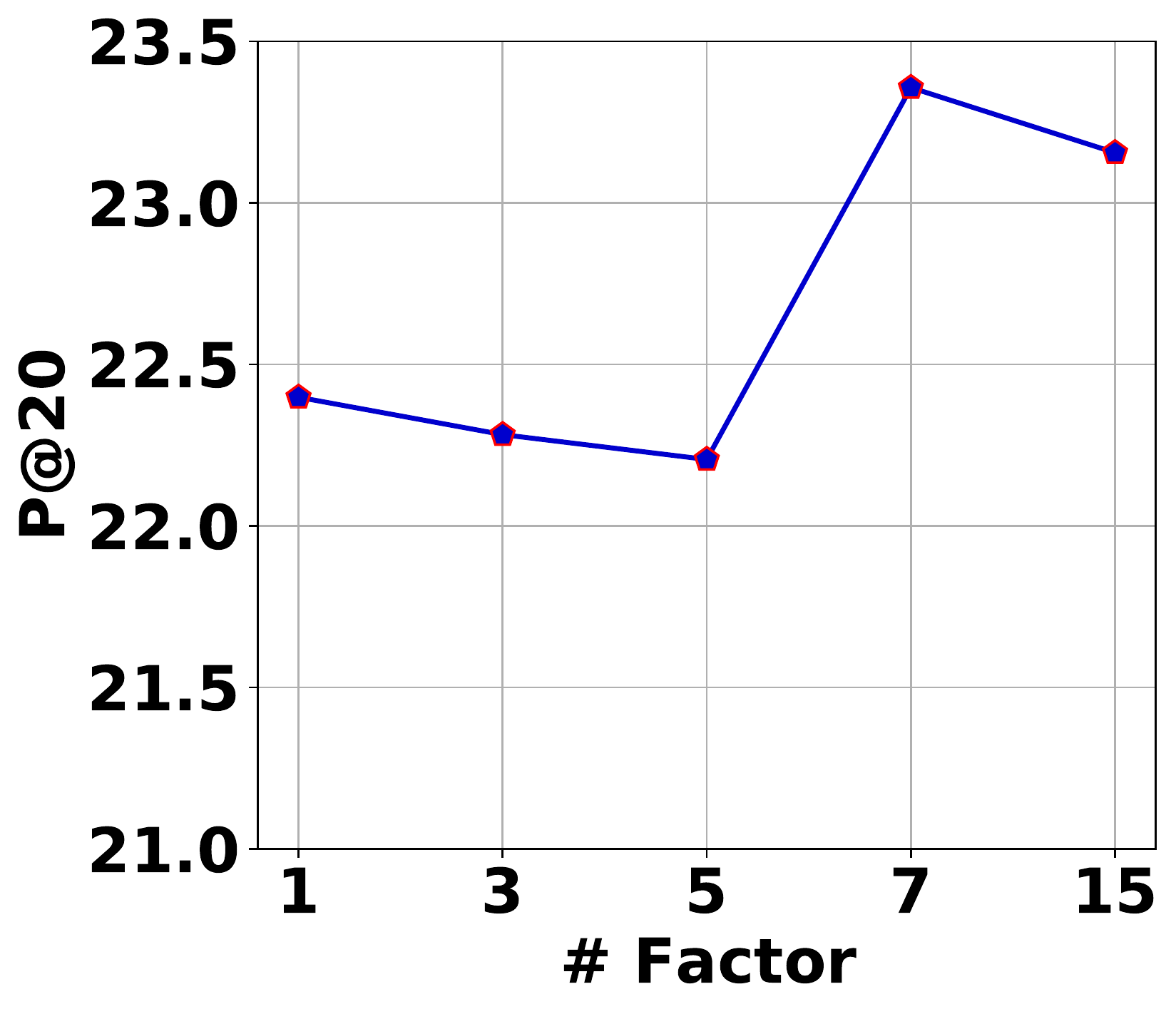}}
	\hspace{25pt}
  \subfloat[MRR@20 on Nowplaying]{\includegraphics[width=0.4\linewidth]{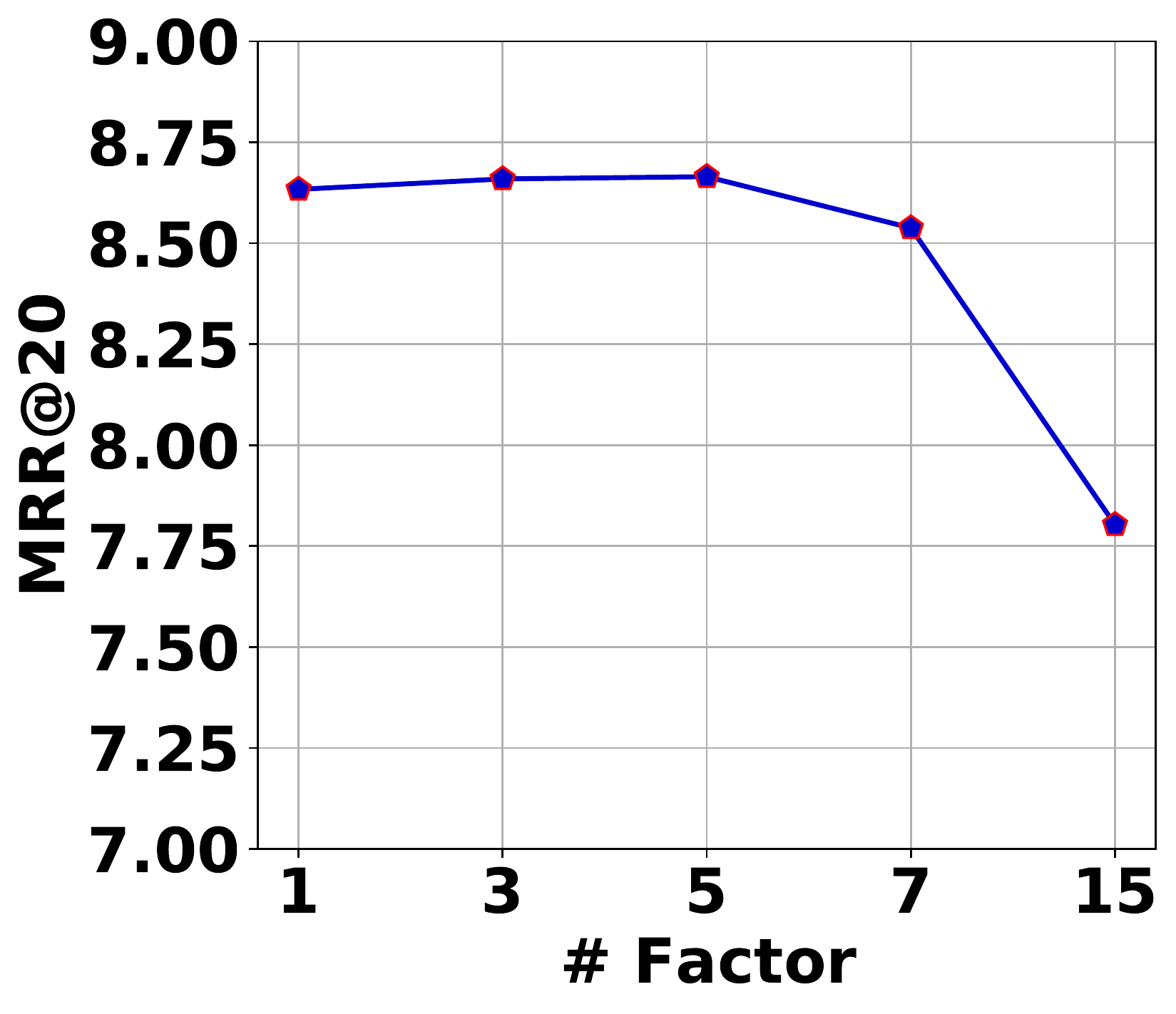}}
	\caption{Impact of factor number ($K$).}
	\vspace{-10pt}
	\label{fig:factor}
\end{figure*}

\subsection{ Influence of key parameters (RQ3)}
\textbf{Influence of disentangled representation ($K$). }To investigate whet-\ her or not TIE-DGNN positively affected by disentangled representation techniques, we study the performance of the model with varying the number of factors ($K$) in the learning process. To be specific, we set item embedding size ($d$) to 128 and 105 for Last.fm and Nowplaying datasets respectively. Then, keep the embedding size unchanged and vary $K$ in two sets of values \{1,2,4,8,16\} (Last.fm) \{1,3,5,7,15\} (Noplaying). The performance comparisons are shown in Fig.\ref{fig:factor}. There are several observations:
\begin{itemize}[leftmargin=1.5em]
  \item{} When $K$ is 1, the performance of our model is relatively worse for both datasets in terms of P@20 and MRR@20 consistently. It indicates that only uniform intent is insufficient to capture a user's specific purpose for different scenarios. This also justifies the rationality of disentangling item embedding for profile users' diverse intents.
  \item With the increase of factors, the performance is generally enhanced till arriving the peak. We find that for different datasets the number of factors corresponding to the peak are varying, e.g.,$K$=4 on Last.fm and $K$=7 on Nowplaying. We think it is due to the different scenarios of datasets, since the user intents are driven by different factors for different scenarios.
  \item When the number of factors is larger than a threshold (i.e., $K$=4 on Last.fm and $K$=7 on Nowplaying), the performance of the model relatively drops. This indicates that too fine-grained intents may result in performance degradation. Thus, selecting the proper $K$ is crucial to disentangling item embeddings in the learning process.
\end{itemize}
  \begin{figure*}[t]
    \centering
    \subfloat{\includegraphics[width=0.42\linewidth]{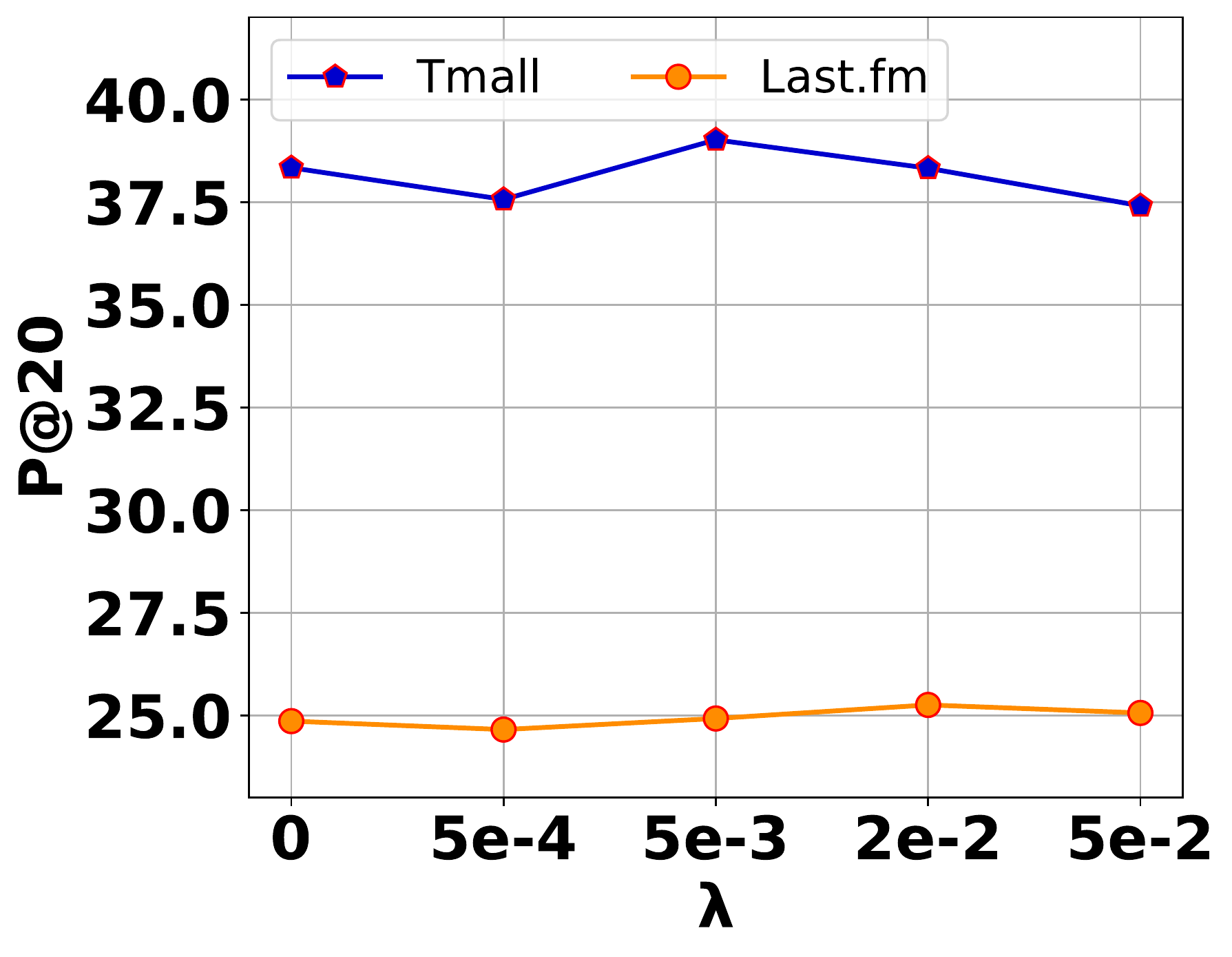}}
    \hspace{25pt}
    \subfloat{\includegraphics[width=0.4\linewidth]{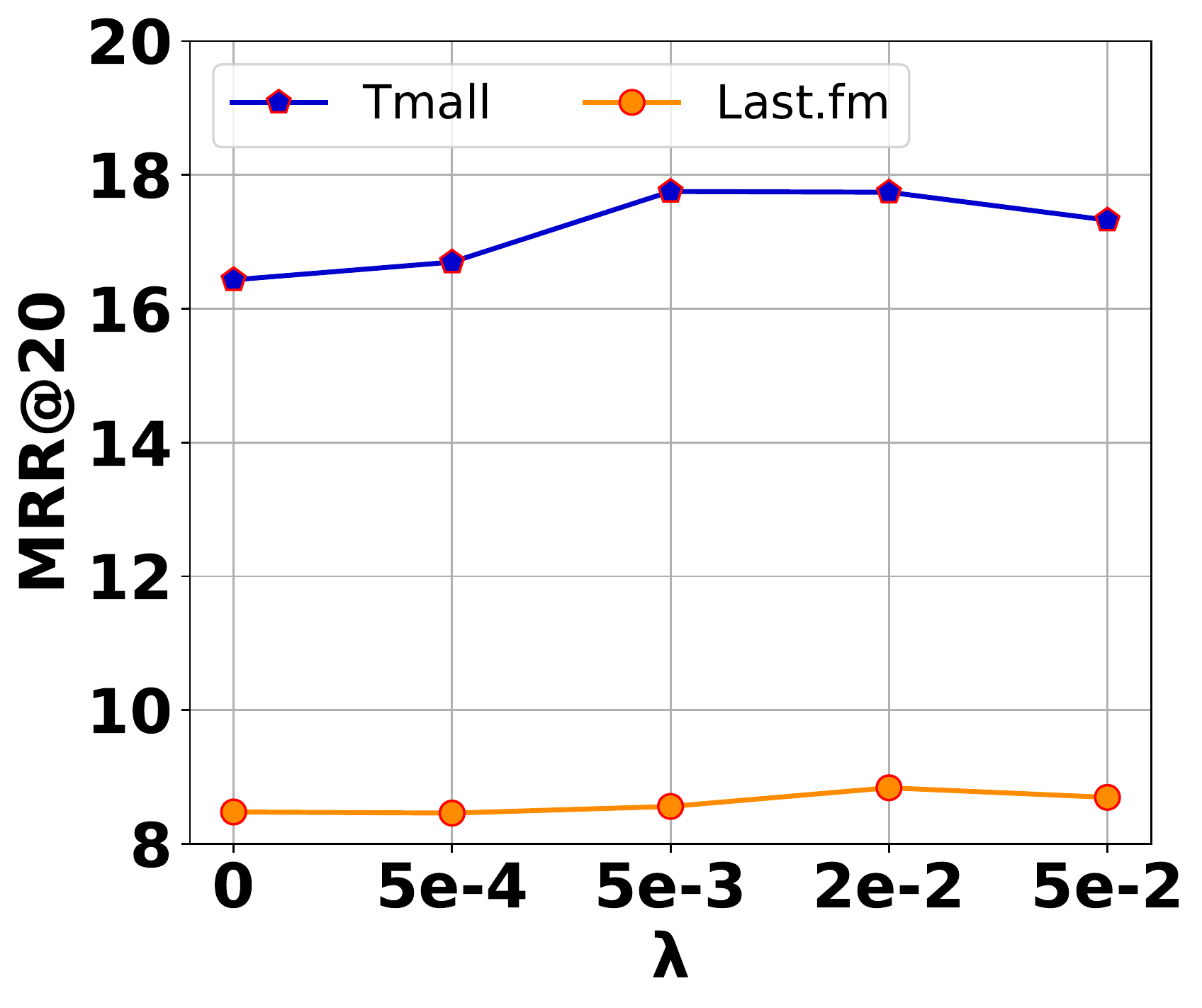}}

    \caption{Impact of the coefficient of contrastive loss ($\lambda$).}
    \vspace{-10pt}
    \label{fig:con}
  \end{figure*}

  \begin{figure*}[t]
    \centering
    \subfloat[P@20 on Last.fm]{\includegraphics[width=0.4\linewidth]{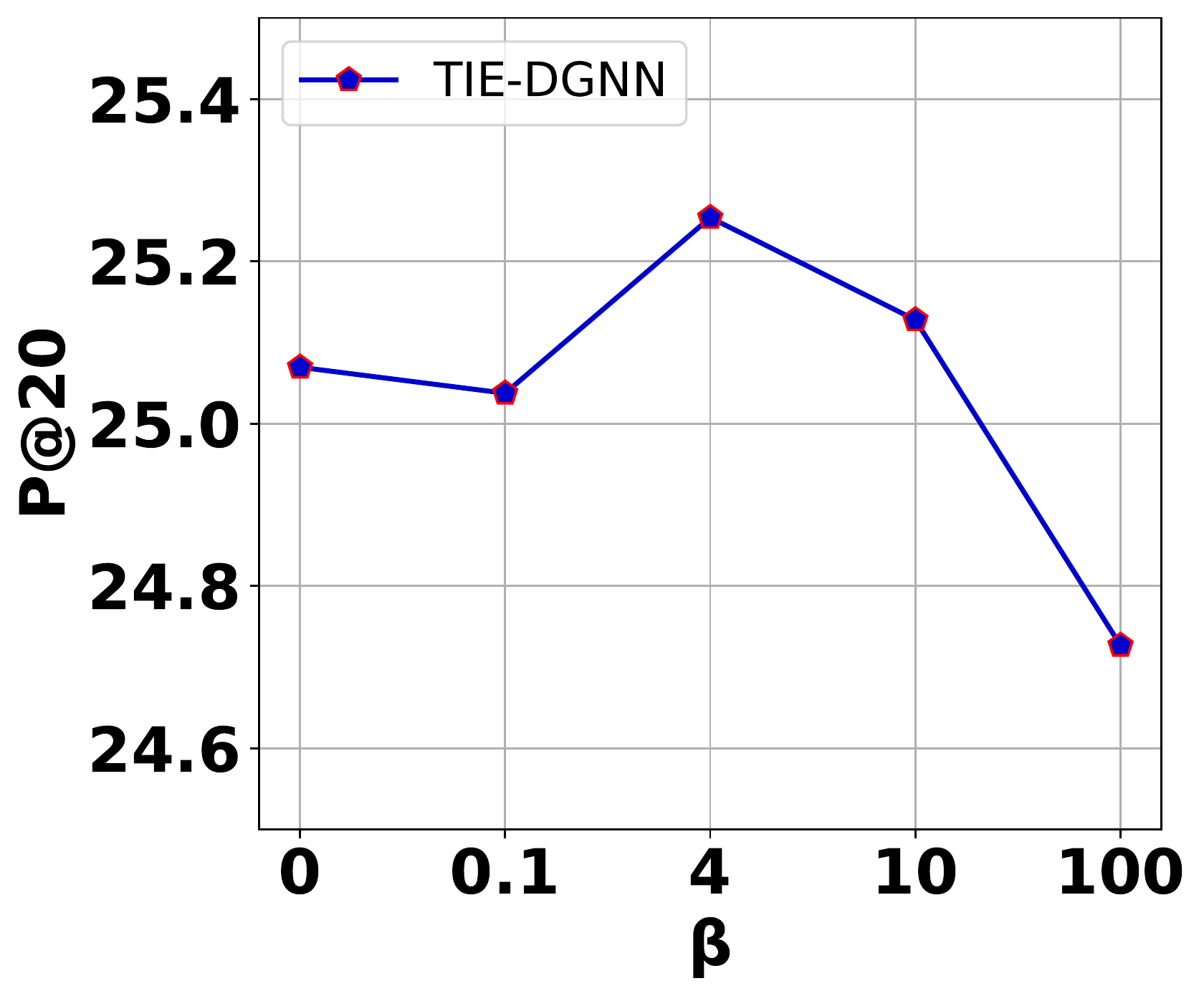}}
    \hspace{25pt}
    \subfloat[P@20 on Tmall]{\includegraphics[width=0.4\linewidth]{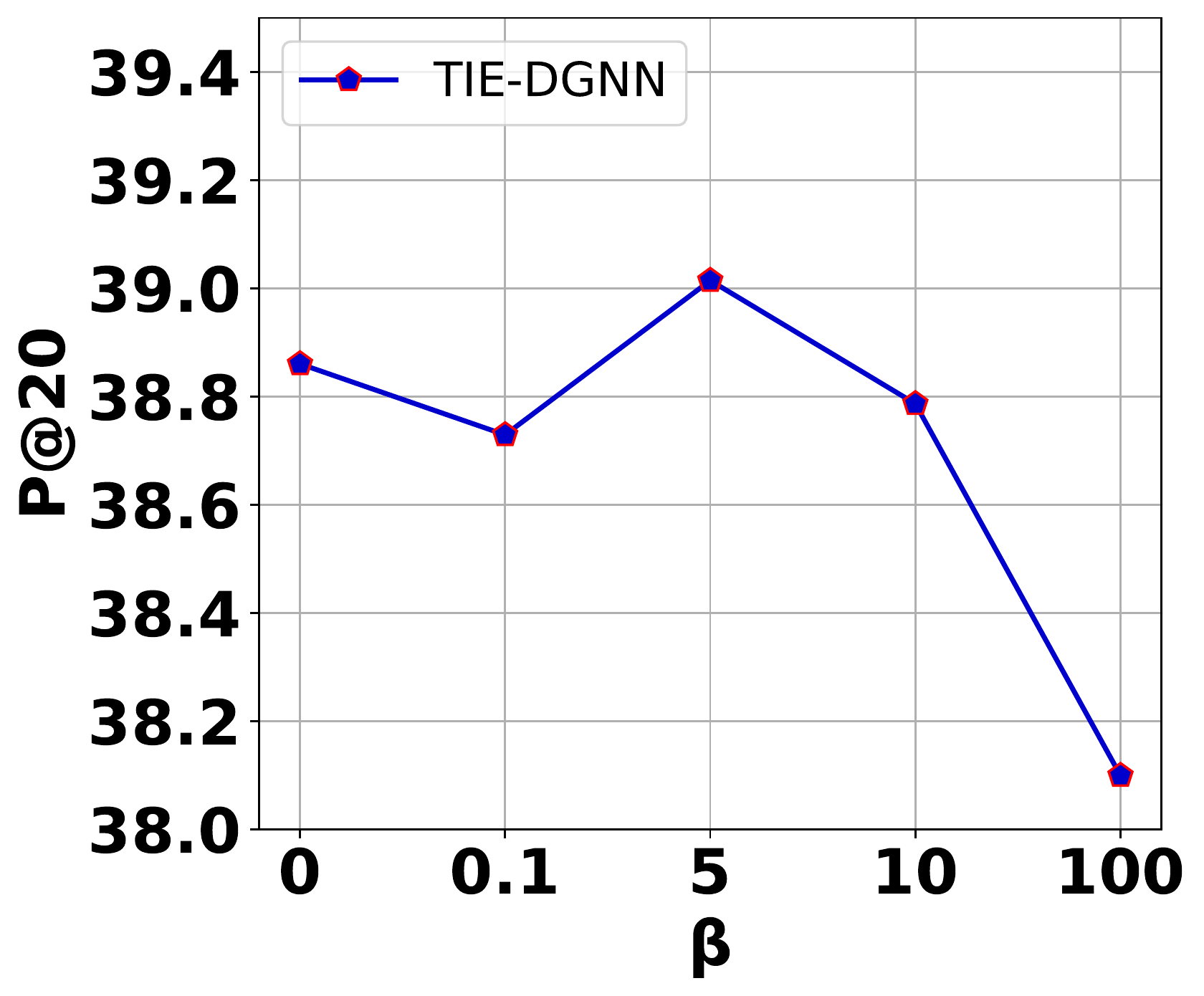}}

    \caption{Impact of the coefficient of disentangled loss ($\beta$).}
    \vspace{-10pt}
    \label{fig:cor}
  \end{figure*}

\textbf{Coefficient of disentangled loss ($\beta$). }To encourage the factor-aware embeddings independence, we apply distance correlation (disentangled loss), \ which is a statistical measure to quantify the level of independence.  The coefficient $\beta$ is a weight of controlling the disentangled loss in the training process. To verify the influence of independence modeling, we tune $\beta$ in a wide range for the best value setting and use P@20 (similar trends are observed based on MRR@20) on Last.fm and Tmall datasets to reflect recommendation performance. The results are shown in Fig.\ref{fig:cor}. It can be seen that both datasets have the same tendencies based on P@20, which rise for a period of fluctuation until reaching a peak and then begin to fall. A properly selected $\beta$ can significantly improve a model's performance and outperform the ones that does not use independent modeling ($\beta$ is 0). It indicates that encouraging factors independent can assist model to obtain better representations in the learning process and thus improve performance. However, when $\beta$ is too large, gradient conflicts between joint loss functions (i.e., contrastive loss function and cross-entropy loss function) lead to the guiding effects of prediction loss on the learning process weakening, and performance suffers as a result.

\textbf{Coefficient of contrastive loss ($\lambda$). }As introduced in Eq.\ref{final}, we apply $\lambda$ to control the magnitude of contrastive loss based on two types of session embeddings. To investigate the influence of contrastive modeling, we tune $\lambda$ in a set of representative values \{0.0005,0.005,0.02,0.05\}. The results are shown in Fig.\ref{fig:con}. When $\lambda$ takes 0.005 and 0.02 for Tmall and Last.fm datasets respectively, the model obtains the best performance in terms of P@20 and MRR@20 consistently. The proper setting of $\lambda$ can significantly enhance the performance of the model. It shows that contrastive learning can enhance the robustness of two types of session embeddings and thus improve the recommendation results. For both datasets, with the increase of $\beta$, the performance of model declines. We think that the reason for the decline is the same as the coefficient $\beta$ of disentangled loss, which is gradient conflicts between joint loss functions.

%% file: 5-conclusion.tex
\section{Conclusion}

In this paper, we propose a novel Transition Information Enhanced Disentangled Graph Neural Network (TIE-DGNN) model for session-based recommendation. Our model highlights the importance of neighboring item transitions from the global context. We first construct a position-aware global graph to model fine-grained transition relationships between items and represent an item as the embeddings of multiple factors to infer the key reason for the transition. Then, we employ global-level disentangling layers to separately learn the factors embeddings of item and train local-level item embeddings via attention mechanism. After obtaining two types of item embeddings including transition information from global and local contexts, we generate factor-aware inter-session embedding and intra-session embedding by taking reversed position information into account, respectively, and use contrastive learning techniques to enhance their robustness. Experimental results show the superiority of our model over the state-of-the-art methods across all the datasets in terms of P@20 and MRR@20. Moreover, further ablation studies verified the validity of different components in our model.